\documentclass[oldversion]{aa}    

\usepackage{graphicx}
\usepackage{txfonts}
\usepackage{natbib}

\bibpunct{(}{)}{;}{a}{}{,}

\begin{document}

\title{Evidence of enhanced star formation efficiency \\ in luminous and ultraluminous infrared galaxies\thanks{Based on observations carried out with the IRAM 30-meter telescope. IRAM is supported by INSU/CNRS (France), MPG (Germany) and IGN (Spain).}}

\author{J.~Graci{\'a}-Carpio\inst{1,2} \and S.~Garc{\'{\i}}a-Burillo\inst{2} \and P.~Planesas\inst{2} \and A.~Fuente\inst{2} \and A.~Usero\inst{2,3}}

\institute{FRACTAL SLNE, Castillo de Belmonte 1, Bloque 5 Bajo A, E-28232 Las Rozas de Madrid, Spain
           \and
           Observatorio Astron{\'o}mico Nacional (OAN), Observatorio de Madrid, Alfonso XII 3, E-28014 Madrid, Spain
           \and
           Centre for Astrophysics Research, University of Hertfordshire, College Lane, AL10 9AB, Hatfield, UK\\
           \email{j.gracia@oan.es, s.gburillo@oan.es, p.planesas@oan.es, a.fuente@oan.es, a.usero@oan.es}}

\date{Received 4 July 2007; accepted 4 December 2007}

\titlerunning{Evidence of enhanced star formation efficiency in luminous and ultraluminous infrared galaxies}

\authorrunning{J.~Graci{\'a}-Carpio et al.}

\abstract{We present new observations made with the IRAM 30m telescope of the J=1--0 and 3--2 lines of HCN and HCO$^{+}$ used to probe the dense molecular gas content in a sample of 17 local luminous and ultraluminous infrared galaxies (LIRGs and ULIRGs). These observations have allowed us to derive an updated version of the power law describing the correlation between the FIR luminosity ($L_{\rm FIR}$) and the HCN(1--0) luminosity ($L'_{\rm HCN(1-0)}$) of local and high-redshift galaxies. We present the first clear observational evidence that the star formation efficiency of the dense gas (SFE$_{\rm dense}$), measured as the $L_{\rm FIR}/L'_{\rm HCN(1-0)}$ ratio, is significantly higher in LIRGs and ULIRGs than in normal galaxies, a result that has also been found recently in high-redshift galaxies. This may imply a statistically significant turn upward in the Kennicutt-Schmidt law derived for the dense gas at $L_{\rm FIR} \geq 10^{11}\,L_{\sun}$. We have used a one-phase Large Velocity Gradient (LVG) radiative transfer code to fit the three independent line ratios derived from our observations. The results of this analysis indicate that the [HCN]$/$[HCO$^{+}$] abundance ratios could be up to one order of magnitude higher than normal in a significant number of LIRGs and ULIRGs of our sample. An overabundance of HCN at high $L_{\rm FIR}$ implies that the reported trend in the $L_{\rm FIR}/L'_{\rm HCN}$ ratio as a function of $L_{\rm FIR}$ would be underestimating a potentially more dramatic change of the SFE$_{\rm dense}$. Results obtained with two-phase LVG models corroborate that the $L'_{\rm HCN(1-0)}$-to-$M_{\rm dense}$ conversion factor must be lowered at high $L_{\rm FIR}$. We discuss the implications of these findings for the use of HCN as a tracer of the dense molecular gas in local and high-redshift luminous infrared galaxies.}


\maketitle

\section{Introduction\label{Introduction}}

The question of how the star formation rate (SFR) in galaxies scales with the density of the star-forming gas is a key problem in extragalactic research. \citet{Schmidt59} first postulated that the SFR per unit volume ($\rho_{\rm SFR}$) should vary as a power law of index $N$ of the gas volume density ($\rho_{\rm gas}$): $\rho_{\rm SFR} \propto \rho_{\rm gas}^{N}$. The translation of the Schmidt law in terms of the corresponding surface densities of SFR ($\Sigma_{\rm SFR}$) and gas ($\Sigma_{\rm gas}$) would imply $\Sigma_{\rm SFR} \propto \Sigma_{\rm gas}^{N}$, if we assume a roughly constant scale-height of the star-forming gas in galaxies. To observationally validate the Schmidt law requires in practice finding fair quantitative tracers of $\Sigma_{\rm SFR}$ and $\Sigma_{\rm gas}$. Above a certain density threshold, determined by large-scale gravitational instability of disks \citep[e.g.,][]{Kennicutt89}, \citet{Kennicutt98} found that the disk-averaged SFRs and gas densities of a sample of $\sim$100 galaxies were well represented by a Schmidt law with an index $N \sim 1.4$, the so called Kennicutt-Schmidt (KS) law. \citet{Kennicutt98} used CO and HI data to derive $\Sigma_{\rm gas}$. \citet[][hereafter GS04a and GS04b]{Gao04a,Gao04b}, using\defcitealias{Gao04a}{GS04a}\defcitealias{Gao04b}{GS04b}a sample of 65 galaxies, derived a similar superlinear correlation between the infrared and the CO(1--0) line luminosities, taken as proxies for the SFR and the total molecular gas content, respectively: $L_{\rm IR} \propto {L'}_{\rm CO(1-0)}^{1.4-1.7}$. Interestingly, they found instead a tight linear correlation (i.e., $N \sim 1$) over 3 decades in $L_{\rm IR}$ between $L_{\rm IR}$ and the luminosity of the HCN(1--0) line $L'_{\rm HCN(1-0)}$, the latter being a tracer of dense molecular gas ($n_{\rm H_{2}} > 10^{4}$\,cm$^{-3}$). More recently \citet{Wu05} have extended a similar correlation to much smaller scales by observing Galactic dense cores in HCN(1--0) emission.

\begin{table*}
\caption{Main properties and IRAM 30m telescope observational results of our sample of LIRGs and ULIRGs. $\theta_{\rm CO}$ is the molecular source size at FWHM derived from interferometric CO observations found in the literature \citep[mainly from][]{Downes98}. Velocity integrated line intensities are given in antenna temperature scale ($T^{*}_{\rm a}$). 1-$\sigma$ uncertainties are derived from the baseline fits. $^{\mathrm{a}}$ Data from \citet{Gracia-Carpio06}.}
\label{Table1}
\centering
\begin{tabular}{@{}lrrrcr r@{}c@{}l@{}c@{}r@{}c@{}l r@{}c@{}l@{}c@{}r@{}c@{}l r@{}c@{}l@{}c@{}r@{}c@{}l r@{}c@{}l@{}c@{}r@{}c@{}l@{}}
\hline
\hline
\noalign{\smallskip}
 Source & \multicolumn{1}{c}{R.A.}    & \multicolumn{1}{c}{Decl.}   & \multicolumn{1}{c}{$D_{L}$} &\multicolumn{1}{c}{$z_{\rm CO}$} & \multicolumn{1}{c}{$\theta_{\rm CO}$} & \multicolumn{7}{c}{$I_{\rm{HCN(1-0)}}$}   & \multicolumn{7}{c}{$I_{\rm{HCN(3-2)}}$}   & \multicolumn{7}{c}{$I_{\rm{HCO}^{+}\rm{(1-0)}}$} & \multicolumn{7}{c}{$I_{\rm{HCO}^{+}\rm{(3-2)}}$} \\ 
        & \multicolumn{1}{c}{(J2000)} & \multicolumn{1}{c}{(J2000)} & $(\rm Mpc)$                 &                                 & \multicolumn{1}{c}{$(\arcsec)$}       & \multicolumn{7}{c}{$\rm (K\,km\,s^{-1})$} & \multicolumn{7}{c}{$\rm (K\,km\,s^{-1})$} & \multicolumn{7}{c}{$\rm (K\,km\,s^{-1})$}        & \multicolumn{7}{c}{$\rm (K\,km\,s^{-1})$}        \\
\noalign{\smallskip}
\hline
\noalign{\smallskip}
 IRAS\,17208--0014 & 17 23 21.90 & $-$00 17 00.1 & 187 \hspace{0.08cm} & 0.04288 & 1.70 & 2&.&19 & \ $\pm$\ \ & 0&.&16$^{\rm a}$ &     4&.&73 & \ $\pm$\ \ & 0&.&43 & 1&.&49 & \ $\pm$\ \ & 0&.&15$^{\rm a}$ &     3&.&06 & \ $\pm$\ \ & 0&.&33$^{\rm a}$ \\
 Mrk 231           & 12 56 14.20 &    56 52 25.9 & 186 \hspace{0.08cm} & 0.04220 & 0.85 & 1&.&39 & \ $\pm$\ \ & 0&.&08           &     1&.&95 & \ $\pm$\ \ & 0&.&22 & 1&.&34 & \ $\pm$\ \ & 0&.&12$^{\rm a}$ &     2&.&04 & \ $\pm$\ \ & 0&.&25$^{\rm a}$ \\
 IRAS\,12112+0305  & 12 13 46.00 &    02 48 41.0 & 331 \hspace{0.08cm} & 0.07291 & 1.00 & 0&.&62 & \ $\pm$\ \ & 0&.&12$^{\rm a}$ & $<$ 1&.&70 &            &  & &   & 0&.&37 & \ $\pm$\ \ & 0&.&09$^{\rm a}$ &      & &   &            &  & &             \\
 Arp 220           & 15 34 57.20 &    23 30 11.5 &  80 \hspace{0.08cm} & 0.01818 & 1.80 & 8&.&16 & \ $\pm$\ \ & 0&.&17           &    18&.&04 & \ $\pm$\ \ & 0&.&51 & 3&.&77 & \ $\pm$\ \ & 0&.&21$^{\rm a}$ &     4&.&57 & \ $\pm$\ \ & 0&.&23           \\
 Mrk 273           & 13 44 42.10 &    55 53 13.1 & 166 \hspace{0.08cm} & 0.03776 & 0.73 & 1&.&11 & \ $\pm$\ \ & 0&.&13           &     3&.&07 & \ $\pm$\ \ & 0&.&61 & 1&.&15 & \ $\pm$\ \ & 0&.&12$^{\rm a}$ &     2&.&90 & \ $\pm$\ \ & 0&.&23$^{\rm a}$ \\
 IRAS\,23365+3604  & 23 39 01.30 &    36 21 10.4 & 280 \hspace{0.08cm} & 0.06438 & 0.95 & 0&.&39 & \ $\pm$\ \ & 0&.&07$^{\rm a}$ & $<$ 0&.&87 &            &  & &   & 0&.&26 & \ $\pm$\ \ & 0&.&06$^{\rm a}$ &      & &   &            &  & &             \\
 UGC 05101         & 09 35 51.60 &    61 21 11.6 & 173 \hspace{0.08cm} & 0.03931 & 3.50 & 1&.&40 & \ $\pm$\ \ & 0&.&14           &      & &   &            &  & &   & 0&.&75 & \ $\pm$\ \ & 0&.&13           &     2&.&13 & \ $\pm$\ \ & 0&.&24           \\
 VII Zw 31         & 05 16 46.70 &    79 40 12.0 & 238 \hspace{0.08cm} & 0.05429 & 2.24 & 0&.&48 & \ $\pm$\ \ & 0&.&07           &      & &   &            &  & &   & 0&.&56 & \ $\pm$\ \ & 0&.&08$^{\rm a}$ &     1&.&06 & \ $\pm$\ \ & 0&.&26           \\
 NGC 6240          & 16 52 58.80 &    02 24 03.8 & 106 \hspace{0.08cm} & 0.02448 & 2.00 & 2&.&18 & \ $\pm$\ \ & 0&.&17           &     9&.&18 & \ $\pm$\ \ & 0&.&68 & 4&.&06 & \ $\pm$\ \ & 0&.&21$^{\rm a}$ &     4&.&24 & \ $\pm$\ \ & 0&.&52$^{\rm a}$ \\
 Arp 55            & 09 15 55.20 &    44 19 54.7 & 176 \hspace{0.08cm} & 0.03984 & 4.50 & 0&.&74 & \ $\pm$\ \ & 0&.&07$^{\rm a}$ &      & &   &            &  & &   & 0&.&69 & \ $\pm$\ \ & 0&.&09$^{\rm a}$ &     0&.&95 & \ $\pm$\ \ & 0&.&23           \\
 Arp 193           & 13 20 35.30 &    34 08 24.6 & 104 \hspace{0.08cm} & 0.02335 & 1.50 & 1&.&04 & \ $\pm$\ \ & 0&.&09           &     1&.&13 & \ $\pm$\ \ & 0&.&21 & 1&.&67 & \ $\pm$\ \ & 0&.&13$^{\rm a}$ &     3&.&11 & \ $\pm$\ \ & 0&.&22$^{\rm a}$ \\
 NGC 695           & 01 51 14.30 &    22 34 56.2 & 136 \hspace{0.08cm} & 0.03245 & 4.00 & 0&.&43 & \ $\pm$\ \ & 0&.&08$^{\rm a}$ &      & &   &            &  & &   & 0&.&62 & \ $\pm$\ \ & 0&.&08$^{\rm a}$ & $<$ 0&.&69 &            &  & &             \\
 Arp 299\,A        & 11 28 33.50 &    58 33 45.3 &  47 \hspace{0.08cm} & 0.01044 & 5.00 & 2&.&04 & \ $\pm$\ \ & 0&.&11$^{\rm a}$ &     1&.&58 & \ $\pm$\ \ & 0&.&38 & 3&.&88 & \ $\pm$\ \ & 0&.&16$^{\rm a}$ &     4&.&72 & \ $\pm$\ \ & 0&.&44$^{\rm a}$ \\
 Arp 299\,B+C      & 11 28 30.80 &    58 33 48.3 &  47 \hspace{0.08cm} & 0.01044 & 7.00 & 1&.&29 & \ $\pm$\ \ & 0&.&09$^{\rm a}$ & $<$ 0&.&72 &            &  & &   & 2&.&09 & \ $\pm$\ \ & 0&.&16$^{\rm a}$ &     1&.&64 & \ $\pm$\ \ & 0&.&24$^{\rm a}$ \\
 NGC 7469          & 23 03 15.60 &    08 52 26.3 &  65 \hspace{0.08cm} & 0.01643 & 4.24 & 1&.&85 & \ $\pm$\ \ & 0&.&09$^{\rm a}$ &     2&.&76 & \ $\pm$\ \ & 0&.&34 & 2&.&70 & \ $\pm$\ \ & 0&.&10$^{\rm a}$ &     2&.&08 & \ $\pm$\ \ & 0&.&42           \\
 Mrk 331           & 23 51 26.80 &    20 35 10.0 &  72 \hspace{0.08cm} & 0.01805 & 4.00 & 1&.&35 & \ $\pm$\ \ & 0&.&10$^{\rm a}$ &     1&.&16 & \ $\pm$\ \ & 0&.&27 & 1&.&74 & \ $\pm$\ \ & 0&.&09$^{\rm a}$ &     2&.&18 & \ $\pm$\ \ & 0&.&41$^{\rm a}$ \\
 NGC 7771          & 23 51 24.90 &    20 06 42.6 &  56 \hspace{0.08cm} & 0.01428 & 4.00 & 3&.&81 & \ $\pm$\ \ & 0&.&13$^{\rm a}$ &     3&.&33 & \ $\pm$\ \ & 0&.&58 & 3&.&67 & \ $\pm$\ \ & 0&.&30$^{\rm a}$ &     1&.&98 & \ $\pm$\ \ & 0&.&37           \\
\noalign{\smallskip}
\hline 
\end{tabular}
\end{table*}

On theoretical grounds it is expected that the power law of the KS law should be close to 1.5. If star formation proceeds due to small-scale gravitational collapse of the gas, an index $N = 1.5$ comes naturally, provided that a constant fraction of the gas forms stars in a free-fall time \citep[e.g.,][]{Larson88}. Alternatively, in a Toomre-stable galaxy disk, an index $N = 1.5$ is expected if a constant fraction of the disk gas forms stars per unit dynamical time-scale, this being determined by the rotation period \citep[e.g.,][]{Elmegreen02}. More recently, \citet{Krumholz05} have advanced a model for star formation regulated by supersonic turbulence where $\rho_{\rm SFR}$ would be a power law of the mean density of the gas, $\overline{\rho}_{\rm gas}$, with an index $N \sim 1.5$. 

Based on the model developed by \citet[][see also discussion in \citealt{Krumholz07a}]{Krumholz05}, \citet{Krumholz07b} have provided a framework describing to what extent the derived correlation between $\rho_{\rm SFR}$ and $\overline{\rho}_{\rm gas}$ would depend on the molecular line tracer used as a proxy for $\overline{\rho}_{\rm gas}$. Line transitions of high effective critical densities ($n_{\rm eff}\sim 10^{4-5}$\,cm$^{-3}$; e.g, like the low-J rotational lines of HCN and HCO$^{+}$) would only trace high-density peaks, and they would be thus insensitive to the bulk of the molecular gas in galaxies. On the contrary, transitions characterized by lower effective critical densities ($n_{\rm eff}\sim 10^{2-3}$\,cm$^{-3}$; e.g., like the low-J rotational lines of CO) would be better tracers of the median density of the bulk of the gas. In this scenario, the different power indexes of SFR so far obtained using either CO or HCN could be explained without violating the universality of the SFR law in galaxies described above. In particular,  \citet{Krumholz07b} argue that the linear correlation between $L_{\rm IR}$ and $L'_{\rm HCN(1-0)}$ found by \citetalias{Gao04b} cannot be taken as firm evidence that the objects detected by the HCN(1--0) line represent a physically distinct star forming unit in molecular gas. 

The use of HCN lines as a \emph{true quantitative} tracer of the dense molecular gas in galaxies can face
difficulties if the excitation conditions and/or the chemical environment of molecular gas depart from normalcy. This can significantly change the conversion factor between the luminosity of HCN lines and the mass of dense molecular gas. From the observational point of view, there is mounting evidence that HCN lines can be \emph{overluminous with respect to the lines of other dense molecular gas tracers}, like HCO$^{+}$, in the circumnuclear disks of Seyferts \citep{Kohno01,Usero04,Kohno05}. A significant percentage of luminous and ultraluminous infrared galaxies (LIRGs and ULIRGs) has also been reported to show high HCN$/$HCO$^{+}$ intensity ratios \citep{Gracia-Carpio06,Imanishi06,Imanishi07}. A similar result has been recently found by \citet{Garcia-Burillo06} in the BAL quasar APM~08279+5255 at $z \sim 4$. The origin of \emph{overluminous} HCN lines, in the terms described above, is still unclear and several theoretical scenarios have been advanced in the literature. X-rays may significantly enhance the abundance of HCN relative to other molecular species like HCO$^{+}$ in enshrouded AGNs, where X-ray Dominated Regions (XDR) can develop \citep{Lepp96}. More recently, \citet{Meijerink05} and \citet{Meijerink07} have proposed that Photon Dominated Regions (PDRs) are more efficient than XDRs in elevating HCN-to-HCO$^{+}$ ratios in starburst galaxies. Hot core-like chemistry in starbursts have also been invoked as a mechanism responsible of enhancing HCN \citep{Lintott06}. Furthermore, instead of being collisionally excited, HCN lines might reflect the pumping by IR photons. These conditions can prevail in the molecular circumnuclear disks around the strong mid infrared sources typically found in AGNs \citep{Aalto95,Garcia-Burillo06,Guelin07,Weiss07,Aalto07a,Aalto07b}. 

The caveats on the use of HCN as the \emph{only} standard tracer of the dense molecular gas in galaxies call for the observation of other molecular species and transitions, in particular in LIRGs and ULIRGs. This question is central to disentangling the different power sources of the huge infrared luminosities of these galaxies. In this paper we present new observations made with the IRAM 30m telescope of the J=1--0 and 3--2 lines of HCN and HCO$^{+}$ used to probe the dense molecular gas content of a sample of 17 LIRGs and ULIRGs. Preliminary results of this work were published by \citet{Gracia-Carpio07a}. These observations, that complement the first HCO$^{+}$ survey published by \citet[][hereafter GC06]{Gracia-Carpio06}\defcitealias{Gracia-Carpio06}{GC06}of LIRGs and ULIRGs, are used to derive a new version of the power law describing the correlation between $L_{\rm FIR}$ and $L'_{\rm HCN(1-0)}$ from normal galaxies ($L_{\rm FIR} < 10^{11}\,L_{\sun}$) to high-$z$ galaxies.  We present the first clear observational evidence that the star formation efficiency of the dense gas, measured as the $L_{\rm FIR}/L'_{\rm HCN(1-0)}$ ratio, is significantly higher in LIRGs and ULIRGs than in normal galaxies. We also find that [HCN]$/$[HCO$^{+}$] abundance ratios could be up to one order of magnitude higher than normal in a significant number of LIRGs and ULIRGs. We discuss the implications of these findings for the use of HCN as a tracer of the dense molecular gas in local and high-redshift luminous infrared galaxies.

\section{Observations\label{Observations}}

\begin{figure*}
\centering
\scalebox{0.63}{\includegraphics{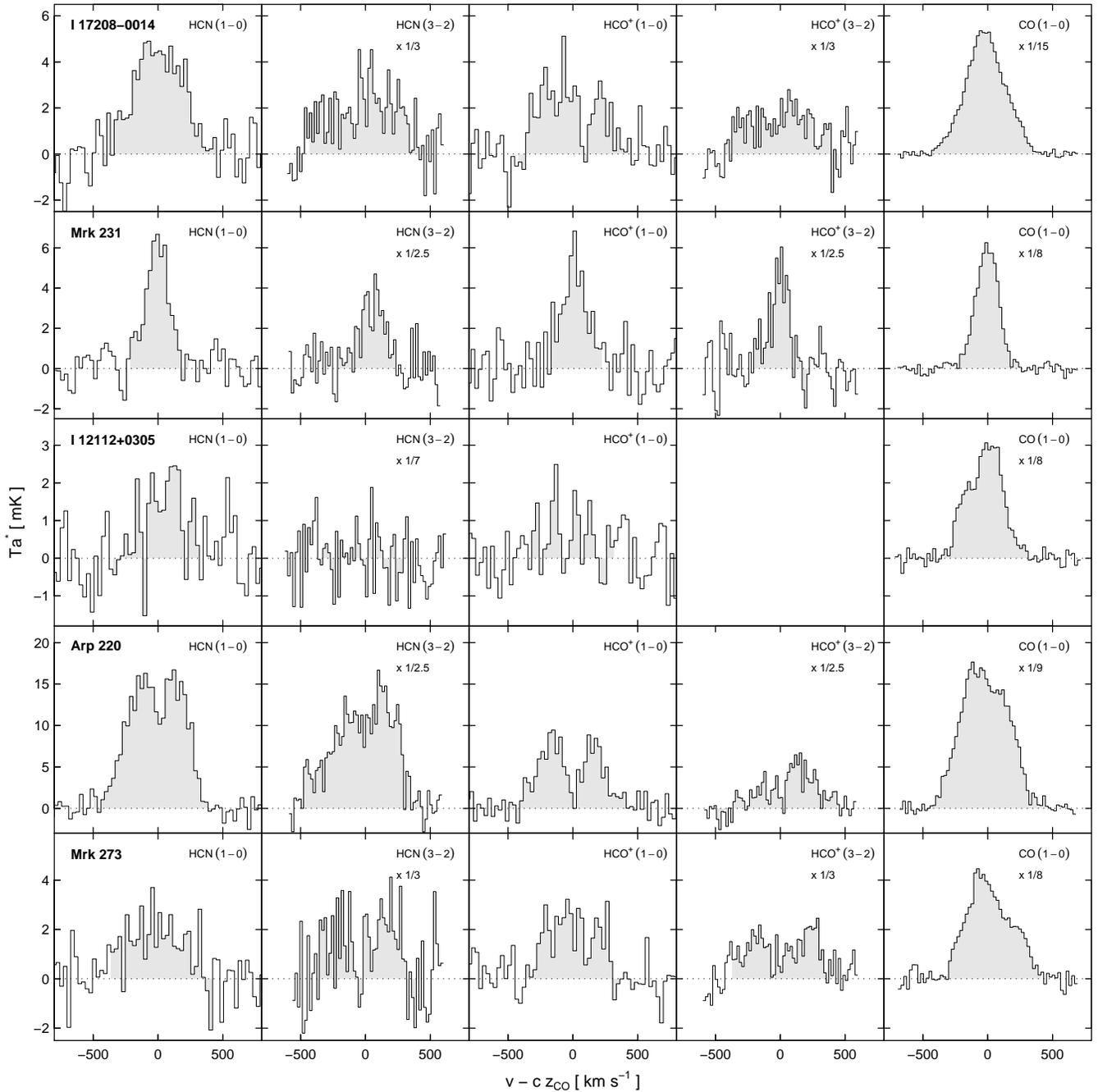}} 
\caption{HCN, HCO$^{+}$ and CO spectra observed with the IRAM 30m telescope. The HCN(3--2), HCO$^{+}$(3--2) and CO(1--0) line intensities have been scaled by the factor indicated in the panel. The velocity windows used to calculate the line areas given in Table~\ref{Table1} are highlighted in grey. Galaxies are ordered with decreasing $L_{\rm FIR}$.}
\label{lines}
\end{figure*}

\begin{figure*}
\centering
\scalebox{0.63}{\includegraphics{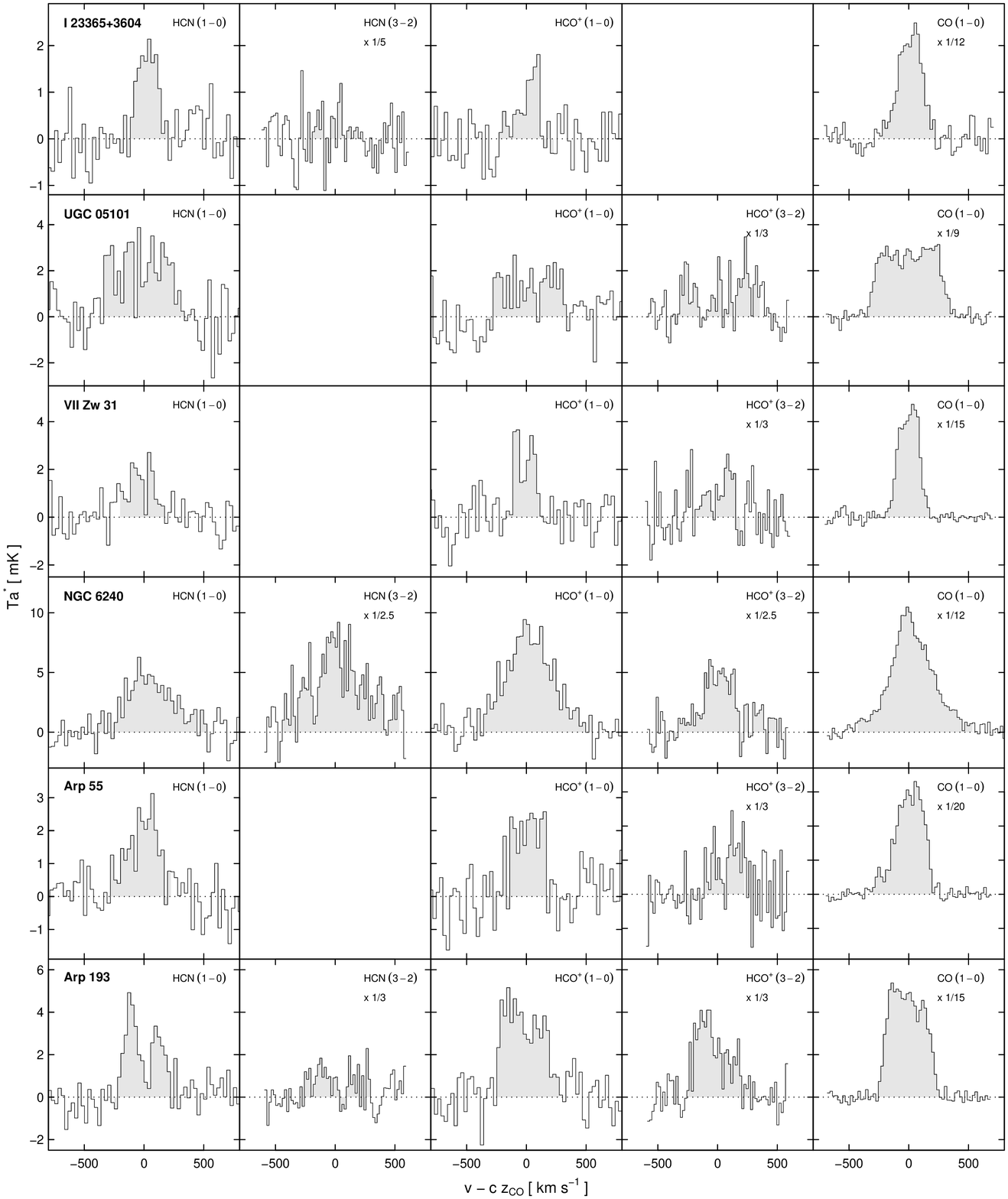}} 
\addtocounter{figure}{-1}
\caption{Continued}
\end{figure*}

\begin{figure*}
\centering
\scalebox{0.63}{\includegraphics{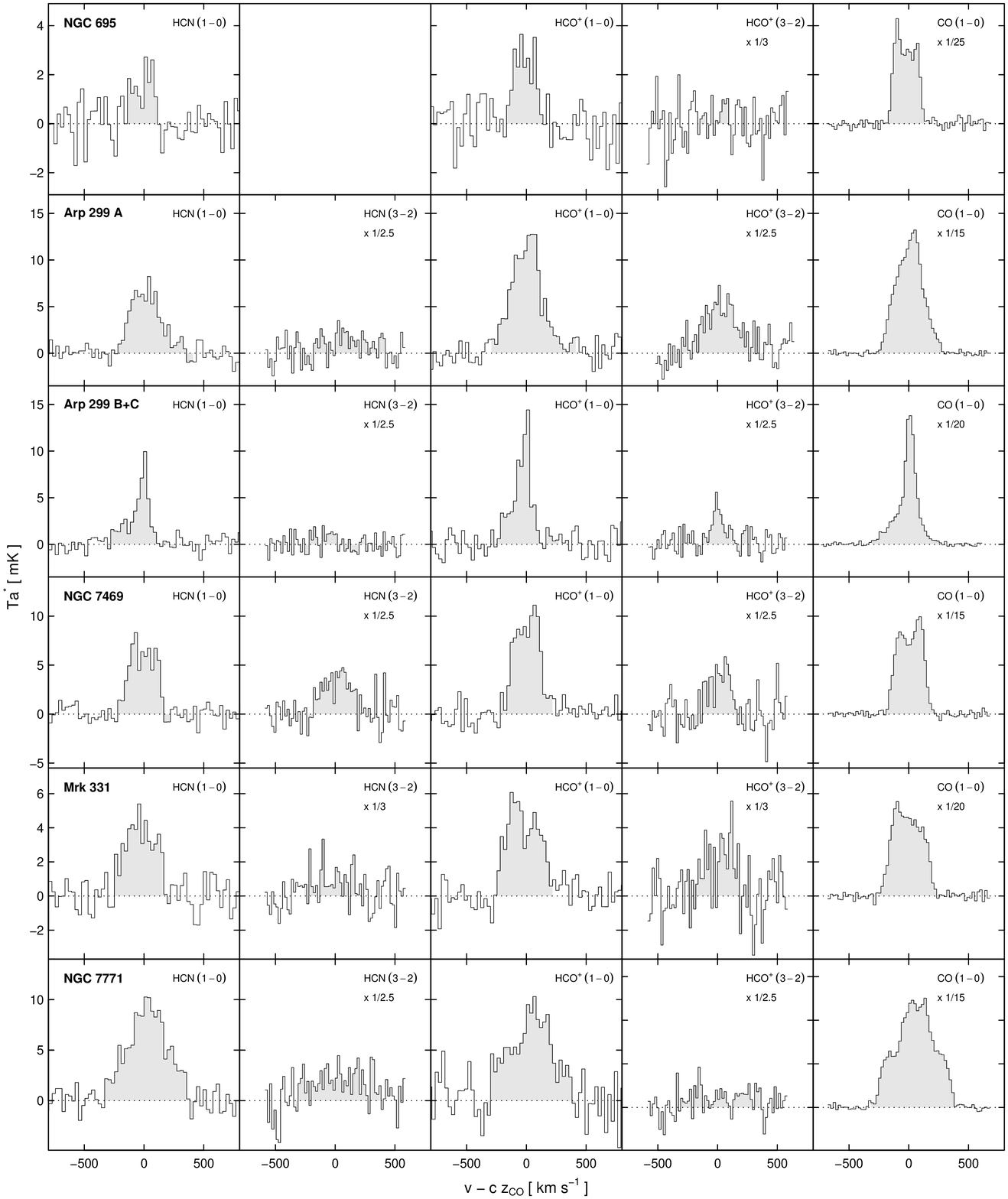}} 
\addtocounter{figure}{-1}
\caption{Continued}
\end{figure*}

\begin{table*}
\caption{Derived luminosities and surface densities of our sample of LIRGs and ULIRGs. $L' = \rm K\,km\,s^{-1}\,pc^{2}$. $^{\mathrm{a}}$ Data from \citetalias{Gracia-Carpio06}, corrected to the different cosmology adopted in this paper and to the finite source size of the molecular gas estimated from interferometric CO observations found in the literature \citep[mainly from][]{Downes98}.}
\label{Table2}
\centering
\begin{tabular}{@{}lrr r@{}c@{}l@{}c@{}r@{}c@{}l r@{}c@{}l@{}c@{}r@{}c@{}l r@{}c@{}l@{}c@{}r@{}c@{}l r@{}c@{}l@{}c@{}r@{}c@{}l @{} rr@{}}
\hline
\hline
\noalign{\smallskip}
 Source & \multicolumn{1}{c}{$L_{\rm{FIR}}$} & \multicolumn{1}{c}{$L_{\rm{IR}}$} & \multicolumn{7}{c}{$L'_{\rm{HCN(1-0)}}$} & \multicolumn{7}{c}{$L'_{\rm{HCN(3-2)}}$} & \multicolumn{7}{c}{$L'_{\rm{HCO}^{+}\rm{(1-0)}}$} & \multicolumn{7}{c}{$L'_{\rm{HCO}^{+}\rm{(3-2)}}$} & \multicolumn{1}{c}{$\Sigma_{\rm SFR}$} & \multicolumn{1}{c}{$\Sigma_{\rm dense}$} \\ 
        & $(10^{11}\,L_{\sun})$              & $(10^{11}\,L_{\sun})$             & \multicolumn{7}{c}{$(10^{8}\,L')$}       & \multicolumn{7}{c}{$(10^{8}\,L')$}       & \multicolumn{7}{c}{$(10^{8}\,L')$}                & \multicolumn{7}{c}{$(10^{8}\,L')$}                & $(M_{\sun}\,{\rm yr^{-1}\,kpc^{-2}})$  & $(M_{\sun}\,{\rm pc^{-2}})$              \\
\noalign{\smallskip}
\hline
\noalign{\smallskip}
 IRAS\,17208--0014 & 18.5 \hspace{0.3cm} & 25.6 \hspace{0.3cm} & 18&.&4  & \ $\pm$\ \ & 1&.&3$^{\rm a}$  &     7&.&2  & \ $\pm$\ \ & 0&.&7  & 12&.&4  & \ $\pm$\ \ & 1&.&2$^{\rm a}$  &     4&.&6  & \ $\pm$\ \ & 0&.&5$^{\rm a}$  & 235 \hspace{0.6cm} & 10400 \hspace{0.3cm} \\
 Mrk 231           & 17.3 \hspace{0.3cm} & 35.0 \hspace{0.3cm} & 11&.&6  & \ $\pm$\ \ & 0&.&7            &     2&.&9  & \ $\pm$\ \ & 0&.&3  & 11&.&0  & \ $\pm$\ \ & 1&.&0$^{\rm a}$  &     3&.&0  & \ $\pm$\ \ & 0&.&4$^{\rm a}$  & 880 \hspace{0.6cm} & 26400 \hspace{0.3cm} \\
 IRAS\,12112+0305  & 15.1 \hspace{0.3cm} & 21.9 \hspace{0.3cm} & 15&.&8  & \ $\pm$\ \ & 3&.&1$^{\rm a}$  & $<$ 7&.&5  &            &  & &   &  9&.&3  & \ $\pm$\ \ & 2&.&3$^{\rm a}$  &      & &   & \      \ \ &  & &             & 220 \hspace{0.6cm} & 10400 \hspace{0.3cm} \\
 Arp 220           & 11.7 \hspace{0.3cm} & 15.5 \hspace{0.3cm} & 12&.&7  & \ $\pm$\ \ & 0&.&3            &     5&.&1  & \ $\pm$\ \ & 0&.&1  &  5&.&8  & \ $\pm$\ \ & 0&.&3$^{\rm a}$  &     1&.&28 & \ $\pm$\ \ & 0&.&06           & 600 \hspace{0.6cm} & 29200 \hspace{0.3cm} \\
 Mrk 273           &  9.9 \hspace{0.3cm} & 14.9 \hspace{0.3cm} &  7&.&4  & \ $\pm$\ \ & 0&.&9            &     3&.&6  & \ $\pm$\ \ & 0&.&7  &  7&.&6  & \ $\pm$\ \ & 0&.&8$^{\rm a}$  &     3&.&4  & \ $\pm$\ \ & 0&.&3$^{\rm a}$  & 620 \hspace{0.6cm} & 20800 \hspace{0.3cm} \\
 IRAS\,23365+3604  &  9.7 \hspace{0.3cm} & 15.2 \hspace{0.3cm} &  7&.&2  & \ $\pm$\ \ & 1&.&3$^{\rm a}$  & $<$ 2&.&8  &            &  & &   &  4&.&8  & \ $\pm$\ \ & 1&.&1$^{\rm a}$  &      & &   & \      \ \ &  & &             & 190 \hspace{0.6cm} &  6340 \hspace{0.3cm} \\
 UGC 05101         &  7.7 \hspace{0.3cm} &  9.9 \hspace{0.3cm} & 10&.&2  & \ $\pm$\ \ & 1&.&0            &      & &   &            &  & &   &  5&.&4  & \ $\pm$\ \ & 0&.&9            &     3&.&0  & \ $\pm$\ \ & 0&.&3            &  18 \hspace{0.6cm} &  1070 \hspace{0.3cm} \\
 VII Zw 31         &  7.4 \hspace{0.3cm} &  9.7 \hspace{0.3cm} &  6&.&5  & \ $\pm$\ \ & 1&.&0            &      & &   &            &  & &   &  7&.&5  & \ $\pm$\ \ & 1&.&1$^{\rm a}$  &     2&.&6  & \ $\pm$\ \ & 0&.&6            &  34 \hspace{0.6cm} &  1330 \hspace{0.3cm} \\
 NGC 6240          &  4.6 \hspace{0.3cm} &  7.1 \hspace{0.3cm} &  6&.&0  & \ $\pm$\ \ & 0&.&5            &     4&.&7  & \ $\pm$\ \ & 0&.&4  & 11&.&0  & \ $\pm$\ \ & 0&.&6$^{\rm a}$  &     2&.&2  & \ $\pm$\ \ & 0&.&3$^{\rm a}$  & 140 \hspace{0.6cm} &  7960 \hspace{0.3cm} \\
 Arp 55            &  4.1 \hspace{0.3cm} &  5.4 \hspace{0.3cm} &  5&.&7  & \ $\pm$\ \ & 0&.&5$^{\rm a}$  &      & &   &            &  & &   &  5&.&2  & \ $\pm$\ \ & 0&.&7$^{\rm a}$  &     1&.&5  & \ $\pm$\ \ & 0&.&4            &   7 \hspace{0.6cm} &   460 \hspace{0.3cm} \\
 Arp 193           &  3.5 \hspace{0.3cm} &  4.8 \hspace{0.3cm} &  2&.&7  & \ $\pm$\ \ & 0&.&2            &     0&.&6  & \ $\pm$\ \ & 0&.&1  &  4&.&4  & \ $\pm$\ \ & 0&.&3$^{\rm a}$  &     1&.&5  & \ $\pm$\ \ & 0&.&1$^{\rm a}$  &  56 \hspace{0.6cm} &  1930 \hspace{0.3cm} \\
 NGC 695           &  3.4 \hspace{0.3cm} &  4.6 \hspace{0.3cm} &  2&.&0  & \ $\pm$\ \ & 0&.&4$^{\rm a}$  &      & &   &            &  & &   &  2&.&8  & \ $\pm$\ \ & 0&.&4$^{\rm a}$  & $<$ 0&.&7  &            &  & &             &  16 \hspace{0.6cm} &   410 \hspace{0.3cm} \\
 Arp 299\,A        &  2.6 \hspace{0.3cm} &  4.0 \hspace{0.3cm} &  1&.&14 & \ $\pm$\ \ & 0&.&06$^{\rm a}$ &     0&.&20 & \ $\pm$\ \ & 0&.&05 &  2&.&15 & \ $\pm$\ \ & 0&.&09$^{\rm a}$ &     0&.&60 & \ $\pm$\ \ & 0&.&06$^{\rm a}$ &  60 \hspace{0.6cm} &  1190 \hspace{0.3cm} \\
 Arp 299\,B+C      &  1.7 \hspace{0.3cm} &  3.1 \hspace{0.3cm} &  0&.&75 & \ $\pm$\ \ & 0&.&05$^{\rm a}$ & $<$ 0&.&11 &            &  & &   &  1&.&19 & \ $\pm$\ \ & 0&.&09$^{\rm a}$ &     0&.&25 & \ $\pm$\ \ & 0&.&04$^{\rm a}$ &  20 \hspace{0.6cm} &   400 \hspace{0.3cm} \\
 NGC 7469          &  2.3 \hspace{0.3cm} &  3.7 \hspace{0.3cm} &  2&.&0  & \ $\pm$\ \ & 0&.&1$^{\rm a}$  &     0&.&62 & \ $\pm$\ \ & 0&.&08 &  2&.&9  & \ $\pm$\ \ & 0&.&1$^{\rm a}$  &     0&.&46 & \ $\pm$\ \ & 0&.&09           &  33 \hspace{0.6cm} &  1270 \hspace{0.3cm} \\
 Mrk 331           &  1.7 \hspace{0.3cm} &  2.6 \hspace{0.3cm} &  1&.&8  & \ $\pm$\ \ & 0&.&1$^{\rm a}$  &     0&.&32 & \ $\pm$\ \ & 0&.&07 &  2&.&2  & \ $\pm$\ \ & 0&.&1$^{\rm a}$  &     0&.&6  & \ $\pm$\ \ & 0&.&1$^{\rm a}$  &  27 \hspace{0.6cm} &  1220 \hspace{0.3cm} \\
 NGC 7771          &  1.6 \hspace{0.3cm} &  2.0 \hspace{0.3cm} &  3&.&0  & \ $\pm$\ \ & 0&.&1$^{\rm a}$  &     0&.&54 & \ $\pm$\ \ & 0&.&09 &  2&.&9  & \ $\pm$\ \ & 0&.&2$^{\rm a}$  &     0&.&33 & \ $\pm$\ \ & 0&.&06           &   7 \hspace{0.6cm} &   545 \hspace{0.3cm} \\
\noalign{\smallskip}
\hline 
\end{tabular}
\end{table*}

The new HCN(1--0), HCO$^{+}$(1--0), HCN(3--2) and HCO$^{+}$(3--2) observations were carried out in five observing runs between December 2005 and November 2006 with the IRAM 30m telescope at Pico de Veleta (Spain). The sample consists of 17 LIRGs and ULIRGs selected to cover homogeneously the $L_{\rm IR}$ range between $10^{11.3}\,L_{\sun}$ and $10^{12.5}\,L_{\sun}$. All galaxies are located at distances larger than 50\,Mpc to be confident that the total emission of the molecular gas can be measured in a single pointing: FWHM(90\,GHz) $\sim 28\arcsec = 7$\,kpc and FWHM(260\,GHz) $\sim 9\arcsec = 2.5$\,kpc, at 50\,Mpc. The 3\,mm and 1\,mm SIS receivers of the 30m telescope were tuned to the redshifted frequencies of the lines. The velocity range covered was $\rm 1800\,km\,s^{-1}$ for the 3\,mm lines and $\rm 1200\,km\,s^{-1}$ for the 1\,mm lines. The wobbler switching mode was used to obtain flat baselines, that was almost always the case for the 3\,mm observations. Only those individual spectra at 1\,mm that showed a flat profile, where an order zero polynomial (i.e., a constant) had to be subtracted, have been retained; others showing a tilted profile were rejected and have not been included in the resulting average spectrum. The velocity ranges used to fit the flat baselines were chosen to be identical to those used for the higher signal-to-noise CO profiles obtained as part of this survey for all the galaxies of our sample.

Typical system temperatures during the observations were $\sim$130\,K at 3\,mm and $\sim$700\,K at 1\,mm. All receivers were used in single side-band mode (SSB), with a high rejection of the image band: $>$12\,dB at 1\,mm and $>$20\,dB at 3\,mm. The latter assures that the calibration accuracy for the bulk of our data is better than 20$\%$. Pointing of the 30m telescope was regularly checked every 2 hours by observing nearby continuum sources; we found an average rms pointing error of 2$\arcsec$--3$\arcsec$ during the observations. When occasionally a larger pointing error was found during an observation, some of the spectra previously taken were dropped, the pointing was corrected, and additional spectra were taken towards the source.

Throughout the paper, velocity-integrated line intensities ($I$) are given in antenna temperature scale, $T_{\rm a}^{*}$. The $T_{\rm a}^{*}$ scale relates to the main beam temperature scale, $T_{\rm mb}$, by the equation $T_{\rm mb} = (F_{\rm eff}/B_{\rm eff}) T_{\rm a}^{*}$, where $F_{\rm eff}$ and $B_{\rm eff}$ are the forward and beam efficiencies of the telescope at a given frequency. For the IRAM 30m telescope $F_{\rm eff}/B_{\rm eff} = 1.22$ (1.96) at 86\,GHz (260\,GHz) and $S/T_{\rm mb} = 4.95$\,Jy\,K$^{-1}$. The velocity window used to calculate $I_{\rm HCO^{+}}$ and $I_{\rm HCN}$ in the two rotational lines has been defined on a case-by-case basis by using the CO(1--0) line profiles. Molecular line luminosities ($L'$) were computed in units of $L' = \rm K\,km\,s^{-1}\,pc^{2}$ according to Equation~1 of \citetalias{Gao04a}, where the K scale here corresponds to the brightness temperature averaged over the size of the source. Luminosity distances have been derived assuming a flat $\rm{\Lambda}$-dominated cosmology described by $H_{0} = 71\,\rm{km\,s}^{-1}\,\rm{Mpc}^{-1}$ and $\rm{\Omega_{m}} = 0.27$ \citep{Spergel03}. Results are summarized in Tables~\ref{Table1} and~\ref{Table2}. All the HCN, HCO$^{+}$ and CO line profiles used in this work are displayed in Fig.~\ref{lines}.

\section{A new HCN/HCO$^{+}$ survey in LIRGs/ULIRGs: comparison with previous data\label{Survey}}

The new data presented in this work represent a significant improvement in the completeness and quality of the available surveys of the dense molecular gas content of LIRGs and ULIRGs. Contrary to the previous survey of \citetalias{Gao04a}, based on a single transition of HCN (the J=1--0 line), the present data set includes another tracer of the dense molecular gas, HCO$^{+}$, and expands the number of observed transitions to two per species. More relevant to the discussion of this paper, and as argued below, the new HCN(1--0) data partly change the results presented in \citetalias{Gao04b}, concerning the constancy of the star formation efficiency in galaxies. 

\begin{figure*}
\centering
\scalebox{0.75}{\includegraphics{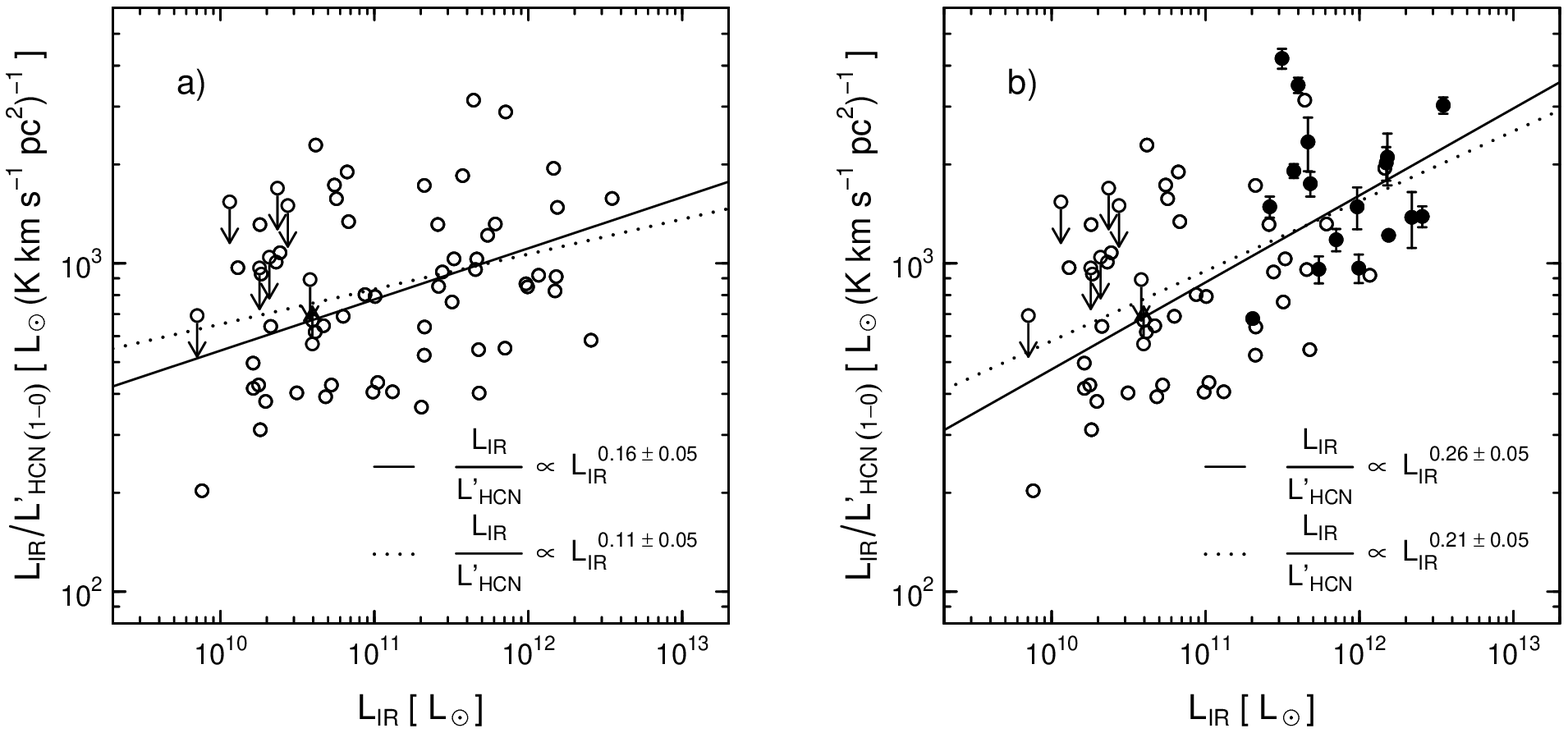}} 
\caption{{\bf (a)} $L_{\rm IR}/L'_{\rm HCN(1-0)}$ luminosity ratio as a function of $L_{\rm IR}$ in the sample of normal galaxies, LIRGs and ULIRGs of \citetalias{Gao04b}. Arrows indicate upper and lower limits to $L'_{\rm HCN(1-0)}$. The solid (dotted) line indicates the orthogonal regression fit calculated for the full sample of galaxies, excluding (including) upper and lower limits. {\bf (b)} Same as {\bf (a)}, but now the filled circles identify the new HCN(1--0) observations of LIRGs and ULIRGs replacing \citetalias{Gao04b}'s old data points.}
\label{comparison}
\end{figure*}

Fig.~\ref{comparison}a represents the $L_{\rm IR}/L'_{\rm HCN(1-0)}$ luminosity ratio as a function of $L_{\rm IR}$ derived from the data published by \citetalias{Gao04b} for a sample of 65 targets including normal galaxies, LIRGs and ULIRGs (hereafter $L_{\rm IR}$ refers to $L_{\rm IR}$(8-1000\,$\mu$m)). This plot is virtually identical to Fig.~2a of \citetalias{Gao04b}; only a few data points have moved in the diagram due to the use of more recently determined values for the distances to the galaxies. The main result of \citetalias{Gao04b}'s paper is echoed by Fig.~\ref{comparison}a: there is no strong trend of the $L_{\rm IR}/L'_{\rm HCN(1-0)}$ luminosity ratio with $L_{\rm IR}$. A power law fit to these data gives $L_{\rm IR}/L'_{\rm HCN(1-0)} \propto L_{\rm IR}^{\ 0.11-0.16}$, depending on whether limits on $L'_{\rm HCN(1-0)}$ are included ($\alpha = 0.11 \pm 0.05$) or excluded ($\alpha = 0.16 \pm 0.05$) in the fit. In terms of average values, the $L_{\rm IR}/L'_{\rm HCN(1-0)}$ luminosity ratio changes from $\sim$900\,$L_{\sun}\,{L'}^{-1}$ for normal galaxies ($L_{\rm IR} < 10^{11}\,L_{\sun}$) to $\sim$1100\,$L_{\sun}\,{L'}^{-1}$ for ULIRGs ($L_{\rm IR} > 8 \times 10^{11}\,L_{\sun}$), i.e., a marginally significant $\sim$20$\%$ increase\footnote{Note that the value of $L_{\rm IR}/L'_{\rm HCN(1-0)} = 740\,L_{\sun}\,{L'}^{-1}$ derived by \citetalias{Gao04b} for normal galaxies, and quoted in Table~2 of their paper, is inconsistent with the data values listed in Table~1 of \citetalias{Gao04b}; based on these data we rather derive $L_{\rm IR}/L'_{\rm HCN(1-0)} \sim 900\,L_{\sun}\,{L'}^{-1}$ for normal galaxies. To ease the comparison with \citetalias{Gao04b} results, in this section we have adopted the same definitions of \emph{normal} galaxy and ULIRG given in their Table~2.}.

Fig.~\ref{comparison}b shows an updated version of the $L_{\rm IR}/L'_{\rm HCN(1-0)}$ luminosity ratio plot as function of $L_{\rm IR}$, where the data of \citetalias{Gao04b} have been replaced by the new HCN(1--0) data obtained with the 30m telescope for 16 LIRGs and ULIRGs. We also include in the plot the new HCN(1--0) data of the ULIRG IRAS\,12112+0305 that was not part of \citetalias{Gao04b}'s sample. Quite noticeably, the inclusion of the new data has a significant effect on creating a tantalizing trend of $L_{\rm IR}/L'_{\rm HCN(1-0)}$ ratio with $L_{\rm IR}$. This trend, hardly evident in Fig.~\ref{comparison}a, can be accommodated by a power law with indexes $\alpha = 0.21 \pm 0.05$ or $\alpha = 0.26 \pm 0.05$, provided that the limits on $L'_{\rm HCN(1-0)}$ are considered or discarded in the fit, respectively. With these new values, we derive a significant increase in the $L_{\rm IR}/L'_{\rm HCN(1-0)}$ luminosity ratio that now goes from $\sim$900\,$L_{\sun}\,{L'}^{-1}$ for normal galaxies to $\sim$1700\,$L_{\sun}\,{L'}^{-1}$ for ULIRGs. This translates into a $\sim$90$\%$ increase in $L_{\rm IR}/L'_{\rm HCN(1-0)}$ over $\sim$2.5 orders of magnitude in $L_{\rm IR}$, i.e., a factor of 4 larger than the corresponding increase derived from \citetalias{Gao04b}'s data.

The reasons explaining the changing picture that results from the comparison of Fig.~\ref{comparison}a and Fig.~\ref{comparison}b, are found on a simple observational fact. A high percentage ($\sim$60$\%$) of the objects newly observed with the 30m telescope show HCN fluxes which are, on average, a factor of 2 lower than those reported by \citetalias{Gao04b}. Objects like Arp~193 are paradigmatic in this respect: the HCN(1--0) flux of Arp~193 in Table~\ref{Table1} is a factor of 5 lower than that reported by \citetalias{Gao04b}. Moreover, in other LIRGs and ULIRGs (like NGC~6240, Mrk~231, Mrk~273 and IRAS~17208--0014), the new HCN(1--0) single-dish fluxes reported in this paper are in much better agreement with the values derived from interferometer maps of the same sources \citep[e.g.,][]{Tacconi99,Nakanishi05,Imanishi06,Imanishi07}. Such an agreement is expected in the case of ULIRGs as the bulk of their molecular gas traced by CO lines and, very likely, of their dense molecular gas content traced by the HCN(1--0) line, is known to be concentrated typically in their central kpc region \citep{Downes98}. In this scenario the chances that interferometer maps may filter out a high percentage of the total HCN(1--0) emission in ULIRGs are scarce. Our results comfortingly fit this picture.

Many of the \textit{conflicting} sources were obtained with the 30m telescope by \citet{Solomon92} in several observing runs during a period going from 1988 to 1991, and were later compiled in \citetalias{Gao04b}. In this context, it is worth noting that the calibration scale of the 30m telescope is more accurate nowadays than two decades ago, owing to the new calibration unit and to the strong rejection of the image side-band of the new generation of receivers installed in the late 1990s (IRAM Newsletter No. 38). In particular, the image side-band rejection used to be less accurately determined around 1990 that it is now, and failure to set the proper sideband rejection could easily overestimate the calibration by up to a factor of 2 (cf. e.g. Kramer 1997, IRAM Report on Calibration of spectral line data). Furthermore, a significant improvement in the quality of the surface of the 30m telescope, due to its re-adjustment in 1997 after holography measurements were carried out, has provided a better knowledge of the efficiencies $F_{\rm eff}$ and $B_{\rm eff}$ both governing the calibration scale of the data, and that are automatically set in the antenna control program since 2000 (IRAM Annual Report 2000). Finally, the pointing accuracy of the telescope is also much better today, due to a better knowledge, monitoring and control of the thermal and mechanical behavior of the antenna, now both taken into account in the pointing model (IRAM Annual Report 2002). Altogether, we conclude that the disagreement between the HCN fluxes given by \citet{Solomon92} and those used in this work can be very likely attributed to occasional calibration problems in the old data.

In the following (Sect.~\ref{SFE}) we explore to what extent the observed trend of $L_{\rm IR}/L'_{\rm HCN(1-0)}$ with $L_{\rm IR}$ can be interpreted as the signature of a change in the star formation laws in galaxies.

\section{Star formation efficiency and Kennicutt-Schmidt laws\label{SFE}}

\subsection{Working hypotheses\label{hypotheses}}

In order to study how the SFR scales with the properties of the dense gas it is necessary to build a large sample of galaxies where $\Sigma_{\rm dense}$ and $\Sigma_{\rm SFR}$ can be both estimated. The objective is to derive an updated version of the KS law for the dense gas using this sample (Sect.~\ref{KS-section}). As a first approach to the problem, we make two assumptions:

\begin{figure}
\centering
\scalebox{0.75}{\includegraphics{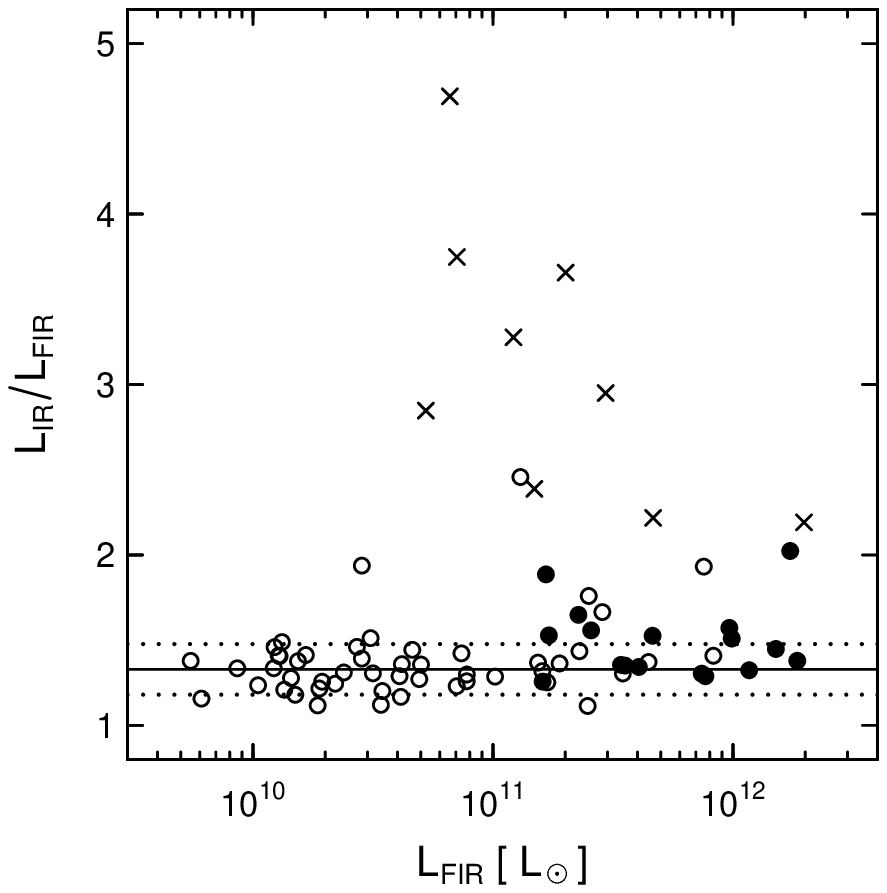}} 
\caption{$L_{\rm IR}/L_{\rm FIR}$ versus $L_{\rm FIR}$ for the sample of normal galaxies, LIRGs and ULIRGs of \citetalias{Gao04b} (open circles), our sample of LIRGs and ULIRGs (filled circles) and a sample of infrared-excess Palomar-Green QSOs \citep[crosses;][]{Evans06}. The solid line indicates the average $L_{\rm IR}/L_{\rm FIR}$ luminosity ratio for normal galaxies ($L_{\rm FIR} < 10^{11}\,L_{\sun}$); the dotted lines show the range of the $\pm$ standard deviation from the mean ($\sim$1.3$\pm$0.15). Infrared-excess Palomar-Green QSOs show the highest $L_{\rm IR}/L_{\rm FIR}$ ratios due to the contribution of the AGN to the total MIR emission of the galaxy.}
\label{LIR-LFIR}
\end{figure}

\begin{figure*}
\centering
\scalebox{0.75}{\includegraphics{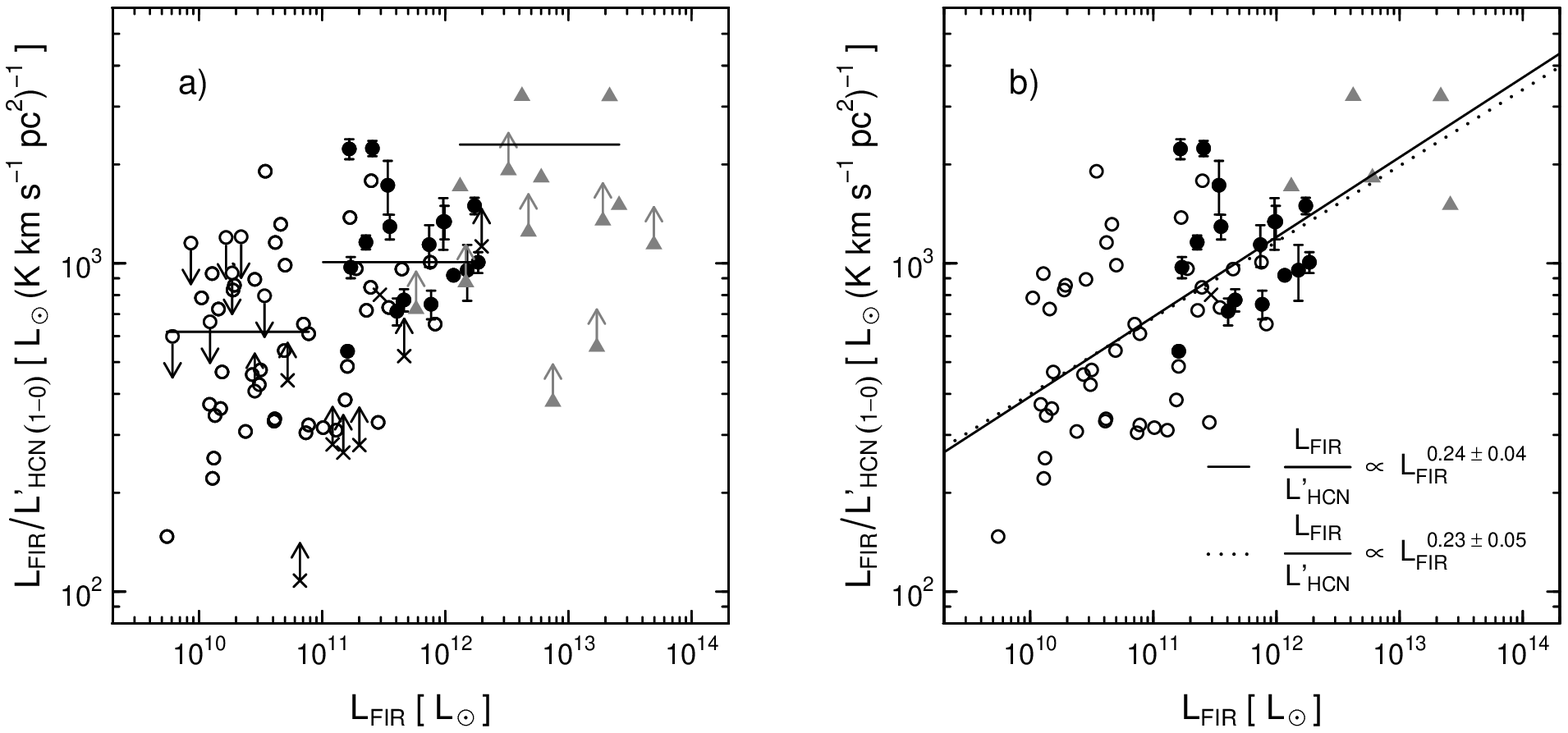}} 
\caption{{\bf (a)} $L_{\rm FIR}/L'_{\rm HCN(1-0)}$ luminosity ratio as a function of $L_{\rm FIR}$ for the full sample of galaxies described in Sec.~\ref{sample}. Symbols are as in Fig.~\ref{LIR-LFIR}. The complete sample of high-$z$ galaxies with available HCN observations is indicated with grey triangles \citep[][and references therein]{Gao07}. Arrows indicate upper and lower limits to $L'_{\rm HCN(1-0)}$. Horizontal lines mark the mean $L_{\rm FIR}/L'_{\rm HCN(1-0)}$ luminosity ratios determined for normal galaxies, IR luminous galaxies and high-$z$ objects. {\bf (b)} Same as {\bf (a)}, but limits have not been represented. The solid (dotted) line indicates the orthogonal regression fit calculated for the full sample of galaxies, including (excluding) high-$z$ objects.} 
\label{SFEplot}
\end{figure*}

\begin{itemize}
  \item First, we consider that $L'_{\rm HCN(1-0)}$ is a good quantitative tracer of the total dense molecular gas mass ($M_{\rm dense}$) of a galaxy independently of $L_{\rm FIR}$, i.e., we implicitly assume that $L'_{\rm HCN(1-0)}$ can be translated into $M_{\rm dense}$ using a \emph{universal} conversion factor. However, the validity of this assumption is revised in Sect.~\ref{XHCN} in the light of the new results presented in this work, pointing to a significant overabundance of HCN in extreme luminous IR galaxies. 
  \item As a second assumption, we also adopt a \emph{universal} conversion factor to derive the SFR of a galaxy from its IR luminosity \citep[e.g.,][]{Kennicutt98}. However, this hypothesis can also be questioned if an AGN significantly contributes to the total IR power of a galaxy ($L_{\rm IR}$(8--1000\,$\mu$m)). In particular, the AGN yield can be fairly large in the MIR range \citep[e.g.,][]{Rowan-Robinson89}. This casts doubts on the results obtained in Sect.~\ref{Survey}, where the reported trend of $L_{\rm IR}/L'_{\rm HCN(1-0)}$ with $L_{\rm IR}$ may simply reflect an increasing AGN contribution to $L_{\rm IR}$, rather than a change in the star formation efficiency.
\end{itemize}

In order to minimize the bias affecting the IR luminosity-to-SFR conversion, we use $L_{\rm FIR}$(40--500\,$\mu$m) to estimate a bolometric IR luminosity which is corrected at best from the AGN contribution. This IR luminosity (denoted $L^{\rm SFR}_{\rm IR}$) is used to estimate the SFR (Sect.~\ref{KS-section}). $L_{\rm FIR}$ is first calculated for each galaxy of the sample fitting its rest frame MIR, FIR and sub-mm spectral energy distribution (SED) to a two grey body model. We then integrate the fit only over the 40--500\,$\mu$m range. This method, that fits the whole IR SED of the galaxy, is more accurate than the standard $L_{\rm FIR}$ determination, that only fits the two IRAS data points at 60\,$\mu$m and 100\,$\mu$m \citep[e.g.][]{Sanders96}. $L_{\rm FIR}$ values are used in Sect.~\ref{SFE-section} to explore the $L_{\rm FIR}$-$L'_{\rm HCN(1-0)}$ correlation in our sample. We then use $L_{\rm FIR}$(40--500\,$\mu$m) to derive $L^{\rm SFR}_{\rm IR}$, re-scaling $L_{\rm FIR}$(40--500\,$\mu$m) by a factor $\sim$1.3. This scaling factor has been determined from the $L_{\rm IR}$(8--1000\,$\mu$m)$/L_{\rm FIR}$(40--500\,$\mu$m) average value derived for the sample of normal galaxies of \citetalias{Gao04b}. The underlying assumption in this calculation is that the AGN contribution is far less substantial in normal galaxies than in ULIRGs or quasars (see Fig.~\ref{LIR-LFIR}). Therefore we can better estimate the bolometric correction that should be applied to derive $L^{\rm SFR}_{\rm IR}$ from $L_{\rm FIR}$(40--500\,$\mu$m) using normal galaxies as templates.

\subsection{Sample\label{sample}}

We have compiled a sample of 88 galaxies with published FIR and HCN observations in the literature. This compilation includes the 17 LIRGs and ULIRGs with new HCN(1--0) observations presented in this paper and the sample of normal galaxies, LIRGs and ULIRGs of \citetalias{Gao04b}. We also include a sample of infrared-excess Palomar-Green QSOs \citep{Evans06} and the complete sample of high-$z$ galaxies with available HCN observations \citep[][and references therein]{Gao07}. 

Because of their redshift, some of these galaxies have not been observed in HCN(1--0), but in HCN(2--1). For these sources $L'_{\rm HCN(1-0)}$ was computed assuming that their rotational line luminosity ratios are $L'_{\rm HCN(2-1)}/L'_{\rm HCN(1-0)} = 0.7$, similar to the mean value measured by \citet{Krips07} in a sample of local galaxies, including a few LIRGs and ULIRGs. For completeness, we have also included in the sample the $z \simeq 4$ quasar APM 08279+5255, observed in the J=5--4 rotational transition of HCN \citep{Wagg05}. To derive its HCN(1--0) luminosity, we adopt $L'_{\rm HCN(5-4)}/L'_{\rm HCN(1-0)} = 0.3$, assuming the physical conditions used in \citet{Garcia-Burillo06} for this source.

\subsection{The star formation efficiency and the $L_{\rm FIR}$-$L'_{\rm HCN(1-0)}$ correlation\label{SFE-section}}

We represent in Fig.~\ref{SFEplot} the $L_{\rm FIR}/L'_{\rm HCN(1-0)}$ luminosity ratio versus $L_{\rm FIR}$ for the full sample of galaxies defined in Sect.~\ref{sample}. Under the assumptions made in Sect.~\ref{hypotheses}, the $L_{\rm FIR}/L'_{\rm HCN(1-0)}$ luminosity ratio can be interpreted as a measure of the star formation efficiency of the dense gas (SFE$_{\rm dense} = {\rm SFR}/M_{\rm dense}$). In Fig.~\ref{SFEplot}a we can see that SFE$_{\rm dense}$ increases with $L_{\rm FIR}$ from normal galaxies to LIRGs \&\ ULIRGs and high-$z$ objects. This result confirms that SFE$_{\rm dense}$ is higher in high-$z$ galaxies compared to normal galaxies \citep{Gracia-Carpio07a,Gao07,Riechers07} and clearly extends this trend for the first time to the luminosity range of local universe LIRGs and ULIRGs. An orthogonal regression fit to the full sample of galaxies, excluding (lower or upper) limits, gives (solid line in Fig.~\ref{SFEplot}b):

\begin{equation}
  \log{\left( \frac{L_{\rm FIR}}{L'_{\rm HCN(1-0)}} \right)} = (0.24 \pm 0.04)\log{L_{\rm FIR}} + (0.17 \mp 0.42) 
\end{equation} 

\begin{equation} 
  \mathrm{or\ } \frac{L_{\rm FIR}}{L'_{\rm HCN(1-0)}} \simeq 1.5\ L_{\rm FIR}^{0.24} 
\end{equation} 

We note that a similar regression fit is obtained if high-$z$ galaxies are not included (dotted line). The latter implies that the regression fit parameters are not at all constrained by high-$z$ galaxies.

\begin{figure}
\centering
\scalebox{0.75}{\includegraphics{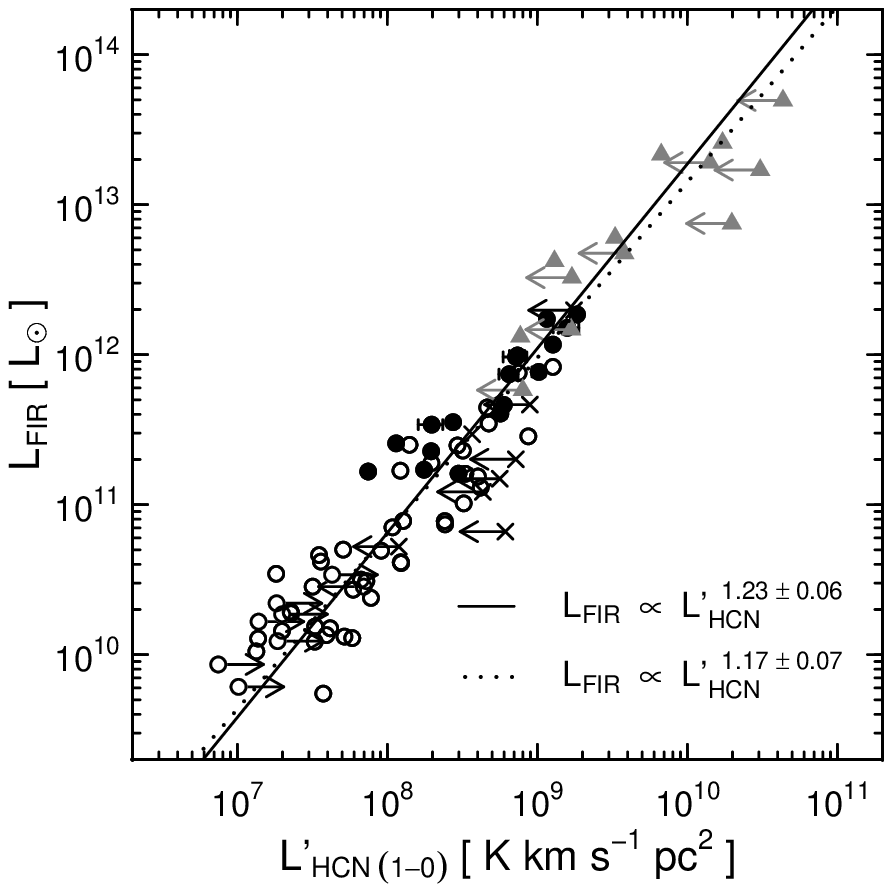}} 
\caption{$L_{\rm FIR}$ versus $L'_{\rm HCN(1-0)}$ using the full sample of galaxies. Arrows indicate upper and lower limits to $L'_{\rm HCN(1-0)}$. The solid (dotted) line represents the orthogonal regression fit to the full sample, including (excluding) high-$z$ objects. As in Fig.~\ref{SFEplot}, limits are not taken into account in the regression fits. Symbols are as in Figs.~\ref{LIR-LFIR} and~\ref{SFEplot}.}
\label{luminosities}
\end{figure}

Another illustration of this result is shown in Fig.~\ref{luminosities}, where we have plotted the $L_{\rm FIR}$-$L'_{\rm HCN(1-0)}$ correlation for the full sample of galaxies. An orthogonal regression fit, excluding limits, gives (solid line in Fig.~\ref{luminosities}):

\begin{equation} 
  \log{L_{\rm FIR}} = (1.23 \pm 0.06)\log{L'_{\rm HCN(1-0)}} + (0.97 \mp 0.46) 
\end{equation}

\begin{equation} 
  \mathrm{or\ } L_{\rm FIR} \simeq 9\ {L'}_{\rm HCN(1-0)}^{1.23} 
\end{equation}

Similarly to the previous case, excluding high-$z$ galaxies from the fit (dotted line) has virtually no effect on the result. In summary, the best-fit correlation found between $L_{\rm FIR}$ and $L'_{\rm HCN(1-0)}$ is seen to be superlinear, contrary to previous claims in the literature \citepalias{Gao04a,Gao04b}. The derived power index is in either case significantly larger than unity: $\sim$1.2.

\subsection{The Kennicutt-Schmidt law\label{KS-section}}

In order to test if a similar superlinear behavior holds for the KS-law of the dense gas, we have determined $\Sigma_{\rm dense}$ and $\Sigma_{\rm SFR}$, using the HCN(1--0) and IR luminosities derived above, and the molecular gas size estimates from published CO interferometer maps, that are available for most of the sources.

We first translate $L^{\rm SFR}_{\rm IR}$ into SFR using the same factor used by \citet{Kennicutt98}\footnote{Note that \citet{Kennicutt98} assumed that the IR luminosity appearing on the right side of his equation, originally $L_{\rm IR}$(8--1000\,$\mu$m), should correspond ideally to radiation coming from star formation reprocessed by dust; as argued in Sect.~\ref{hypotheses}, $L^{\rm SFR}_{\rm IR}$ values derived in the present work are a more educated guess of the SFR than $L_{\rm IR}$(8--1000\,$\mu$m), however.}: 

\begin{equation} 
  {\rm SFR}\ [M_{\sun}\,{\rm yr^{-1}}] = 1.7 \times 10^{-10}\,L^{\rm SFR}_{\rm IR}\ [L_{\sun}] 
\end{equation} 

Similarly, we derive $M_{\rm dense}$ from $L'_{\rm HCN(1-0)}$ assuming the conversion factor of \citetalias{Gao04a}: 

\begin{equation}
  M_{\rm dense}\ [M_{\sun}] = 10\,L'_{\rm HCN(1-0)}\ [\rm K\,km\,s^{-1}\,pc^{2}] 
\end{equation}

\begin{figure*}
\centering
\scalebox{0.75}{\includegraphics{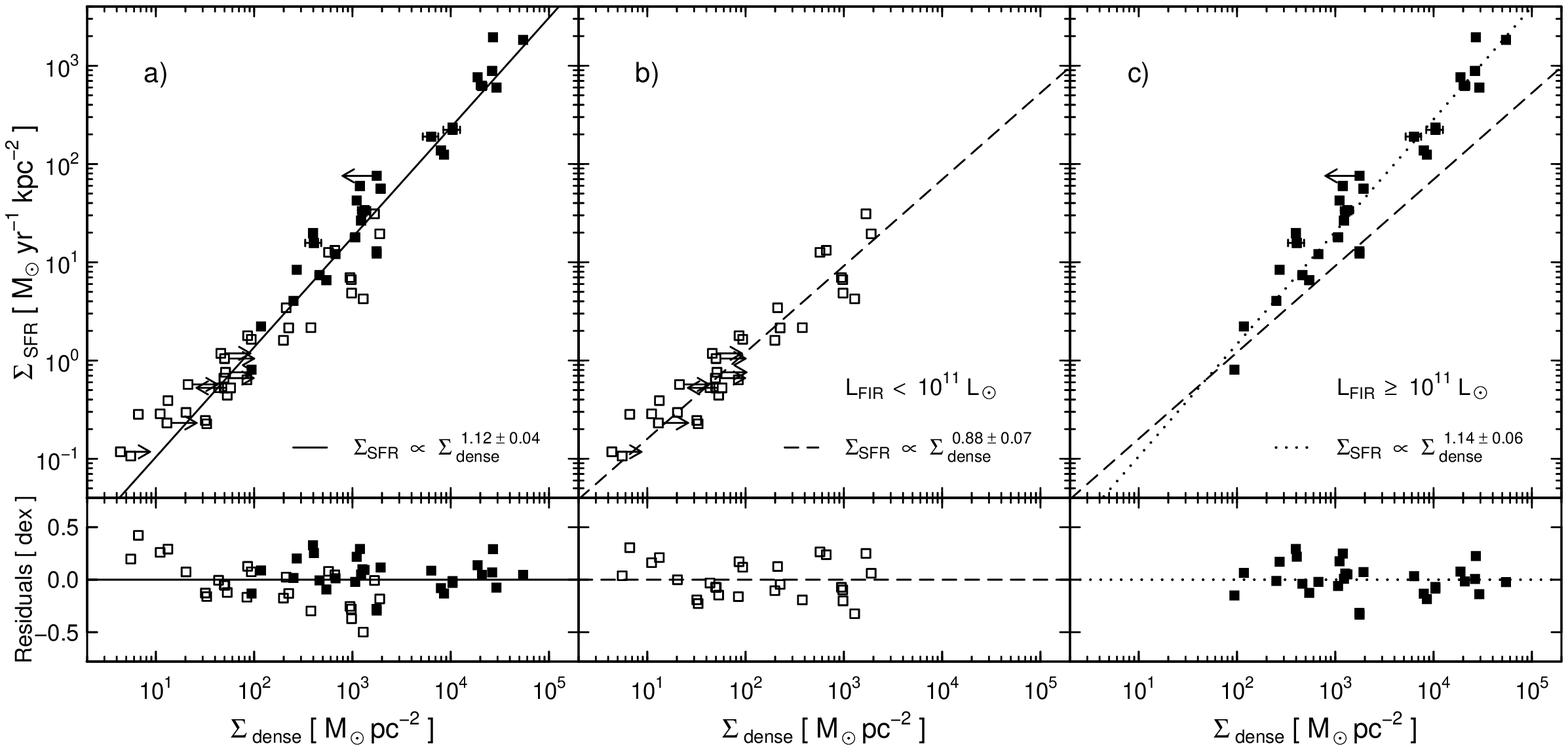}} 
\caption{{\bf (a)} Surface density of star formation rate, $\Sigma_{\rm SFR}$, against surface density of the dense molecular gas mass as traced by the HCN(1--0) line, $\Sigma_{\rm dense}$, for normal (open squares) and IR luminous galaxies (filled squares) for which the molecular gas size is available. Arrows indicate upper and lower limits to $\Sigma_{\rm dense}$. The solid line is the orthogonal regression fit to the full sample of galaxies. In {\bf (b)} and {\bf (c)} we separate between normal
and IR luminous galaxies, respectively. The dashed line is the orthogonal regression fit to galaxies with $L_{\rm FIR} < 10^{11} L_{\sun}$ and the dotted line is a similar fit to galaxies with $L_{\rm FIR} > 10^{11} L_{\sun}$, including one Palomar-Green QSO and five high-$z$ galaxies. Upper and lower limits are not considered in the regression fits. Lower panels show the residuals of the fits in logarithmical units.}
\label{KSlaw}
\end{figure*}

Finally, to obtain $\Sigma_{\rm SFR}$ and $\Sigma_{\rm dense}$ we normalize SFR and $M_{\rm dense}$ using the size of the molecular gas distribution of the sources. In Fig.~\ref{KSlaw}a we have represented the derived $\Sigma_{\rm SFR}$ and $\Sigma_{\rm dense}$ for the full sample of galaxies. $\Sigma_{\rm SFR}$ and $\Sigma_{\rm dense}$ follow a very tight correlation over more than 4 orders of magnitude in $\Sigma_{\rm dense}$. An orthogonal regression fit to the data results in a KS-law of the dense molecular gas with a power index $N = 1.12 \pm 0.04$:

\begin{equation}
  \log{\Sigma_{\rm SFR}} = (1.12 \pm 0.04)\log{\Sigma_{\rm dense}} + (-2.10 \mp 0.12)
\end{equation}

\begin{equation}
  \mathrm{or\ } \Sigma_{\rm SFR} \simeq 0.008\ \Sigma_{\rm dense}^{1.12}
\end{equation}

The fit of the KS-law to the whole sample is still significantly nonlinear ($N > 1$). A close examination to the residuals from this global fit indicates that the population of normal galaxies would be better described by a shallower index power law. As illustrated in the lower panel of Fig.~\ref{KSlaw}a, residuals for galaxies with $L_{\rm FIR} < 10^{11}\,L_{\sun}$ show a systematic trend with $\Sigma_{\rm dense}$ from $0.5$\,dex at $10\,M_{\sun}$\,pc$^{-2}$ to $-0.5$\,dex at $10^3\,M_{\sun}$\,pc$^{-2}$. To investigate if there is a turn upward in the KS law at $L_{\rm FIR} \geq 10^{11}\,L_{\sun}$, we split the sample into normal ($L_{\rm FIR} < 10^{11}\,L_{\sun}$) and IR luminous galaxies ($L_{\rm FIR} > 10^{11}\,L_{\sun}$) and made a two-function power law fit to the whole sample. The result of this fit, displayed in Figs.~\ref{KSlaw}b and \ref{KSlaw}c, indicates that, although the two galaxy populations follow very well defined KS power laws over 3 decades in $\Sigma_{\rm SFR}$ and $\Sigma_{\rm dense}$, their characteristic power indexes are substantially different: $N = 0.88 \pm 0.07$ for normal galaxies, while $N = 1.14 \pm 0.06$ for IR luminous galaxies. The extrapolation to higher $\Sigma_{\rm dense}$ of the KS-law fitting normal galaxies falls short of explaining the typical values of $\Sigma_{\rm SFR}$ in IR luminous galaxies by up to a factor $\sim$6--7 (see Fig.~\ref{KSlaw}c). 

In order to evaluate the significance of the two-function power law fit versus that of a single power law, $\chi^{2}$ values have been derived, resulting in an improvement when the two-function power law fit is adopted: the $\chi^{2}$ value decreases from 1.9 to 1.4. We thus propose that the relation between star formation and the properties of the dense molecular gas in galaxies can be better described by two KS-laws:

\begin{equation}
  \log{\Sigma_{\rm SFR}} = (0.88 \pm 0.07)\log{\Sigma_{\rm dense}} + (-1.68 \mp 0.15)
\end{equation}

\begin{equation} 
  \mathrm{or\ } \Sigma_{\rm SFR} \simeq 0.02\ \Sigma_{\rm dense}^{0.88}
\end{equation}

\noindent for galaxies with $L_{\rm FIR} < 10^{11}\,L_{\sun}$, and:

\begin{equation}
  \log{\Sigma_{\rm SFR}} = (1.14 \pm 0.06)\log{\Sigma_{\rm dense}} + (-2.12 \mp 0.21)
\end{equation}

\begin{equation} 
  \mathrm{or\ } \Sigma_{\rm SFR} \simeq 0.008\ \Sigma_{\rm dense}^{1.14}
\end{equation}

\noindent for local and high-$z$ IR luminous galaxies with $L_{\rm FIR} \geq 10^{11}\,L_{\sun}$. 

In the previous analysis the CO distribution size is used to evaluate the surface densities, because it is available for far more objects than the dense gas distribution sampled with HCN. Any bias due to an overestimation of the molecular gas size corresponding to the dense gas, here taken from CO, should have a negligible effect on the reported tilt of the regression fits for the two families of galaxies. The underlying reason is that both parameters, $\Sigma_{\rm SFR}$ and $\Sigma_{\rm dense}$, are affected to the same extent by this bias. Should a residual effect exist, correcting for it would even reinforce the magnitude of the tilt, as in low luminosity objects the CO distribution tends to be more extended than the dense gas distribution whereas for high luminosity objects, although limited to some cases, there is observational evidence that both distributions are more similar in extent.

In the following sections we discuss the potential bias that we may have introduced in estimating $\Sigma_{\rm SFR}$ and $\Sigma_{\rm dense}$ from $L_{\rm FIR}$ and $L'_{\rm HCN(1-0)}$ according to the assumptions of Sect.~\ref{hypotheses}, and how they may have affected the results presented in this section.

\section{Excitation and chemistry of dense molecular gas in IR luminous galaxies\label{XHCN}}

\citetalias{Gracia-Carpio06} analyzed the data issued from the first HCO$^{+}$ survey of LIRGs and ULIRGs, finding significant evidence that the HCN(1--0)$/$HCO$^{+}$(1--0) luminosity ratio increases with $L_{\rm FIR}$ in infrared luminous galaxies ($L_{\rm FIR} > 10^{11}\,L_{\sun}$; see Fig.~2a of \citetalias{Gracia-Carpio06}'s paper). Fig.~\ref{chemistry}a shows an updated version of the same plot based on the new HCN(1--0) data presented in this paper. The inclusion of this new data set corroborates the existence of a statistically significant trend of the HCN(1--0)$/$HCO$^{+}$(1--0) luminosity ratio with $L_{\rm FIR}$ for $L_{\rm FIR} > 10^{11}\,L_{\sun}$. As argued in Sect.~\ref{Survey}, the greatly improved quality of the new HCN data used to build up Fig.~\ref{chemistry}a gives a higher significance to this result compared to our previous finding \citepalias{Gracia-Carpio06}. Of particular note is the recent independent confirmation of a similar result by \citet{Imanishi07}, who studied a sample of infrared luminous galaxies using the Nobeyama millimeter array. 

Taken at face value, the reality of the trend shown in Fig.~\ref{chemistry}a suggests that the overall excitation and/or the chemical properties of the dense molecular gas change on average with $L_{\rm FIR}$. The inclusion into this analysis of the new J=3--2 line data of HCN and HCO$^{+}$, obtained for 10 galaxies in our sample, can further constrain the physical parameters ($n_{\rm H_{2}}$ and $T_{\rm k}$) and chemical abundance ratios ([HCN]$/$[HCO$^{+}$]) describing the average properties of the dense molecular gas in infrared luminous galaxies. This characterization can be achieved through a radiative transfer modelling of the line ratios.

\subsection{One-phase Large Velocity Gradient (LVG) models\label{one-phase}}

\begin{figure}
\centering
\scalebox{0.75}{\includegraphics{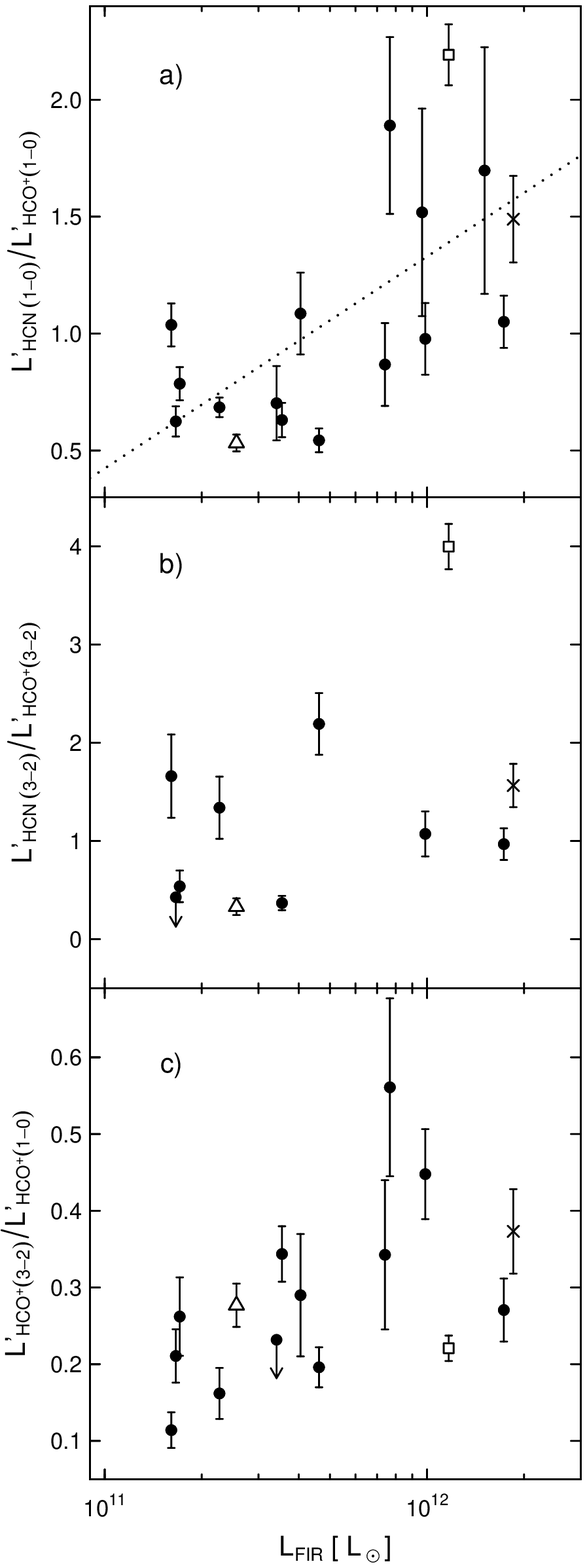}} 
\caption{{\bf (a)} HCN(1--0)$/$HCO$^{+}$(1--0) luminosity ratio as a function of $L_{\rm FIR}$ in our sample of LIRGs and ULIRGs. The dashed line, that represents a linear regression fit to the data points, shows the trend. {\bf (b)} and {\bf (c)} Same as {\bf (a)} but for the HCN(3--2)$/$HCO$^{+}$(3--2) and HCO$^{+}$(3--2)$/$HCO$^{+}$(1--0) luminosity ratios, respectively. The positions of Arp 299\,A (open triangle), IRAS\,17208--0014 (cross) and Arp~220 (open square) are highlighted.}
\label{chemistry}
\end{figure}

Without further observational constraints at hand, the simplest approach consists of using a one-phase Large Velocity Gradient (LVG) scheme to fit the three independent intensity ratios ($R_{i}$) which are derived from the line intensities measured in our survey. To evaluate the goodness of the fits we have applied a standard $\chi^{2}$-test to the obtained solutions. We estimate the $\chi^{2}$ of the 3-parameter fit from the expression:

\begin{equation}
  \chi^{2} = \sum_{i=1}^{3} \frac{(R_{i} - R^{\rm \ model}_{i})^2}{\sigma_{i}^{2}}
\end{equation}

\noindent where $R^{\rm \ model}_{i}$ are the best-fit model predictions for $R_{i}$, and $\sigma_{i}$ are the estimated uncertainties on $R_{i}$; values of $\chi^{2} < 3$ help to identify good fits. We have chosen to fit the following $R_{i}$ ratios: HCN(1--0)$/$HCO$^{+}$(1--0), HCN(3--2)$/$HCO$^{+}$(3--2) and HCO$^{+}$(3--2)$/$HCO$^{+}$(1--0). These ratios, listed in Table~\ref{Obs-ratios}, are displayed in Figs.~\ref{chemistry}a-to-c as a function of $L_{\rm FIR}$ for the infrared luminous galaxies of our sample. Being a one-phase scheme, we have to assume that the beam filling factor of the emission is common for all lines. The LVG filling factors implied by best-fit solutions have been confronted on a case-by-case basis with the values estimated from interferometer CO maps that exist for all the galaxies of our sample (Table~\ref{Table1}). This helps to discard LVG solutions implying beam filling factors ($\eta_{\rm fill}$) for HCN and HCO$^{+}$ lines that exceed those derived from high spatial resolution CO maps.

\begin{table}
\caption{Observed molecular line ratios.}
\label{Obs-ratios}
\centering
\begin{tabular}{@{}l c c c@{}}
\hline
\hline
\noalign{\smallskip}
 Source & $\frac{\rm HCN(1-0)}{\rm HCO^{+}(1-0)}$ & $\frac{\rm HCN(3-2)}{\rm HCO^{+}(3-2)}$ & $\frac{\rm HCO^{+}(3-2)}{\rm HCO^{+}(1-0)}$ \\ 
\noalign{\smallskip}
\hline
\noalign{\smallskip}
 IRAS\,17208--0014 & 1.49 $\pm$ 0.19 & 1.56 $\pm$ 0.22 & 0.37 $\pm$ 0.06 \\
 Mrk 231           & 1.05 $\pm$ 0.11 & 0.97 $\pm$ 0.16 & 0.27 $\pm$ 0.04 \\
 IRAS\,12112+0305  & 1.70 $\pm$ 0.53 &                 &                 \\
 Arp 220           & 2.19 $\pm$ 0.13 & 4.00 $\pm$ 0.23 & 0.22 $\pm$ 0.02 \\
 Mrk 273           & 0.98 $\pm$ 0.15 & 1.07 $\pm$ 0.23 & 0.45 $\pm$ 0.06 \\
 IRAS\,23365+3604  & 1.52 $\pm$ 0.44 &                 &                 \\
 UGC 05101         & 1.89 $\pm$ 0.38 &                 & 0.56 $\pm$ 0.12 \\
 VII Zw 31         & 0.87 $\pm$ 0.18 &                 & 0.34 $\pm$ 0.10 \\
 NGC 6240          & 0.54 $\pm$ 0.05 & 2.19 $\pm$ 0.31 & 0.20 $\pm$ 0.03 \\
 Arp 55            & 1.09 $\pm$ 0.17 &                 & 0.29 $\pm$ 0.08 \\
 Arp 193           & 0.63 $\pm$ 0.07 & 0.37 $\pm$ 0.07 & 0.34 $\pm$ 0.04 \\
 NGC 695           & 0.70 $\pm$ 0.16 &                 & $<$0.29         \\
 Arp 299\,A        & 0.53 $\pm$ 0.04 & 0.33 $\pm$ 0.09 & 0.28 $\pm$ 0.03 \\
 Arp 299\,B+C      & 0.62 $\pm$ 0.06 & $<$0.52         & 0.21 $\pm$ 0.03 \\
 NGC 7469          & 0.68 $\pm$ 0.04 & 1.34 $\pm$ 0.32 & 0.16 $\pm$ 0.03 \\
 Mrk 331           & 0.79 $\pm$ 0.07 & 0.54 $\pm$ 0.16 & 0.26 $\pm$ 0.05 \\
 NGC 7771          & 1.04 $\pm$ 0.09 & 1.66 $\pm$ 0.42 & 0.11 $\pm$ 0.02 \\
\hline 
\end{tabular}
\end{table}

\begin{figure*}
\centering
\scalebox{0.75}{\includegraphics{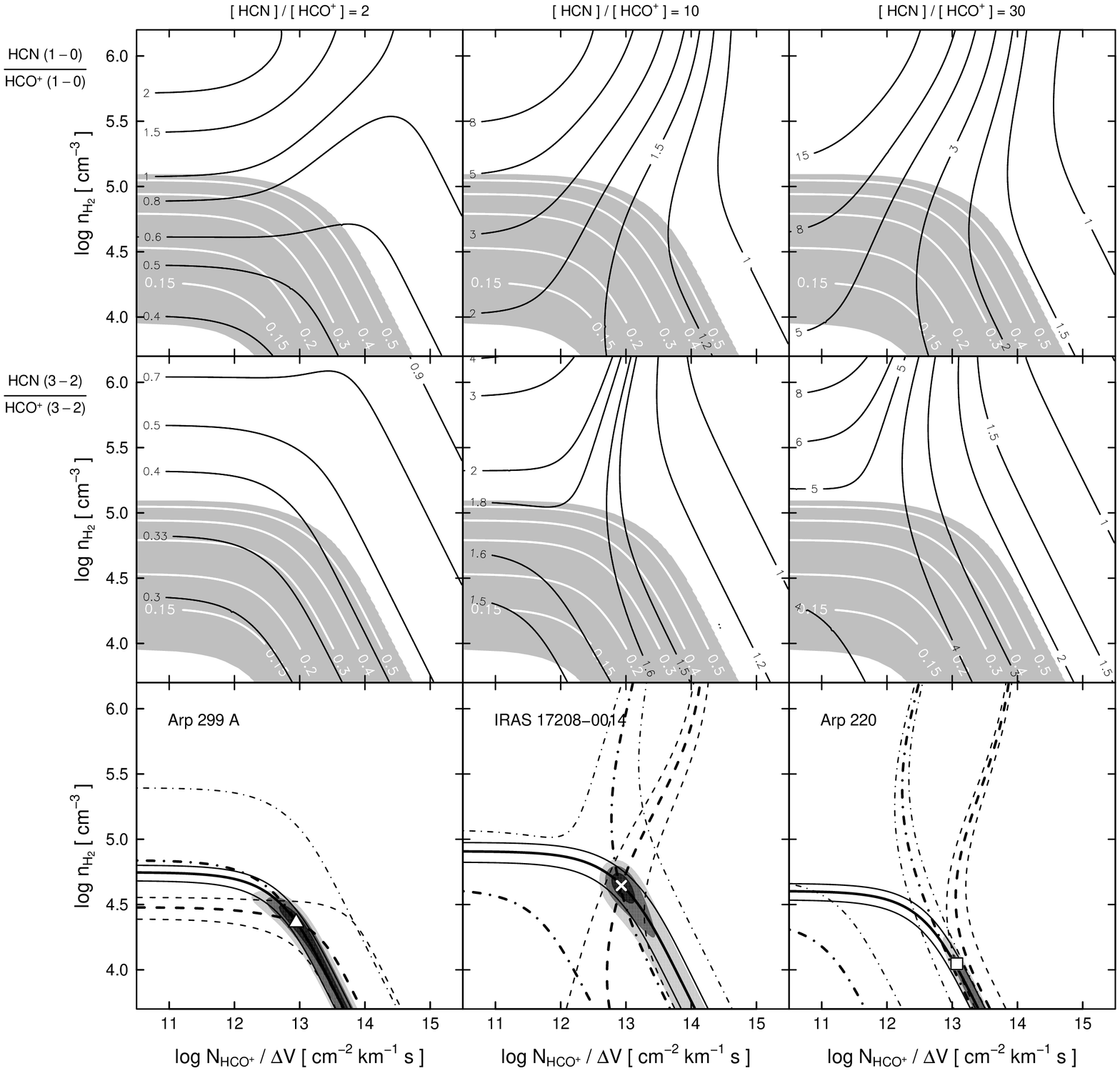}} 
\caption{Results of the one-phase LVG calculations for the HCN(1--0)$/$HCO$^{+}$(1--0) (black contours, upper panels) and HCN(3--2)$/$HCO$^{+}$(3--2) line ratios (black contours, middle panels) as a function of the HCO$^{+}$ column density per velocity interval ($N_{\rm HCO^{+}}/\Delta V$), the H$_{2}$ volume density ($n_{\rm H_{2}}$) and the [HCN]$/$[HCO$^{+}$] relative abundance ($= 2$, 10 and 30 from left to right). The grey area indicates the range of HCO$^{+}$(3--2)$/$HCO$^{+}$(1--0) line ratios (white contours, upper and middle panels) observed in our sample of LIRGs and ULIRGs. In the bottom panels we show the one-phase LVG best fit solution of three representative sources: Arp 299\,A, IRAS\,17208--0014 and Arp~220. Filled, dashed, and dot-dashed contours represent, respectively, the regions where the predicted HCO$^{+}$(3--2)$/$HCO$^{+}$(1--0), HCN(1--0)$/$HCO$^{+}$(1--0) and HCN(3--2)$/$HCO$^{+}$(3--2) line ratios are equal to the observational values given in Table~\ref{Obs-ratios}. Thinner lines indicate one-$\sigma$ deviations from these ratios. Grey areas correspond to regions where $\chi^{2} \leq 3$, 1.5 and 0.5, from light grey to black. The kinetic temperature was fixed to 60\,K in all the calculations.}
\label{LVG}
\end{figure*}

\begin{figure}
\centering
\scalebox{0.75}{\includegraphics{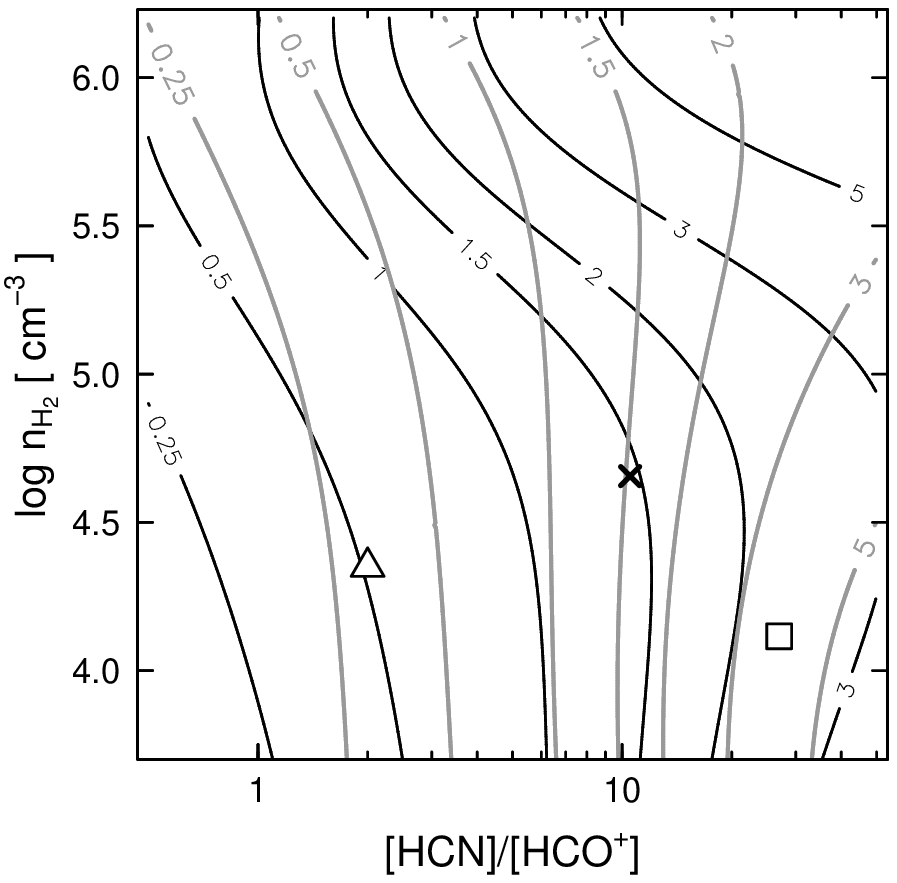}} 
\caption{Results of the one-phase LVG calculations for the HCN(1--0)$/$HCO$^{+}$(1--0) (black contours) and HCN(3--2)$/$HCO$^{+}$(3--2) line ratios (grey contours) as a function of the [HCN]/[HCO$^{+}$] abundance ratio and the H$_{2}$ volume density ($n_{\rm H_{2}}$). The HCO$^{+}$ column density per velocity interval ($N_{\rm HCO^{+}}/\Delta V$) and the kinetic temperature ($T_{\rm K}$) have been fixed to $10^{13}$\,cm$^{-2}$\,km$^{-1}$\,s and 60\,K, respectively, to illustrate the range of solutions for [HCN]/[HCO$^{+}$]. Symbols indicate the best fit solutions of the three representative sources discussed in Sect.~\ref{one-phase}: Arp 299\,A (open triangle), IRAS\,17208--0014 (cross) and Arp~220 (open square).}
\label{LVG2}
\end{figure}

In summary, we find that, for any common set of physical parameters ($n_{\rm H_{2}}$ and $T_{\rm k}$) simultaneously fitting all the line ratios, we require [HCN]$/$[HCO$^{+}$] abundance ratios to be $>$5 for a significant number of LIRGs and ULIRGs in our sample (Table~\ref{LVG-results}). This is illustrated in the upper and middle panels of Fig.~\ref{LVG}, that show the LVG model predictions of the three considered line ratios as a function of $n_{\rm H_{2}}$, the HCO$^{+}$ column density per velocity interval ($N_{\rm HCO^{+}}/\Delta V$) and the [HCN]$/$[HCO$^{+}$] abundance ratio. Within the one-phase scenario, HCN$/$HCO$^{+}$ luminosity ratios $>$1 in the J=1--0 and 3--2 lines cannot be explained with [HCN]$/$[HCO$^{+}$] $\leq 2$ for the range of HCO$^{+}$(3--2)$/$HCO$^{+}$(1--0) luminosity ratios observed in our sample of LIRGs and ULIRGs, requiring higher abundances of HCN relative to HCO$^{+}$ (see also Fig.~\ref{LVG2}). This result is mostly independent of the value of $T_{\rm k}$ adopted in the models (values in the range $T_{\rm k} = 20$ to 100\,K have been explored in our search for the best fit solution) and it is a direct consequence of the critical densities of the rotational lines of HCN being a factor of $\sim$6 higher than those of HCO$^{+}$. Quite interestingly, the case for overabundant HCN is more compelling at high $L_{\rm FIR}$.

The bottom panels of Fig.~\ref{LVG} illustrate the best fit solutions obtained for $T_{\rm k} = 60$\,K in three galaxies: Arp~299\,A, IRAS\,17208--0014 and Arp 220. These are fair representatives of the three different categories of solutions identified in our sample of LIRGs and ULIRGs (Table~\ref{LVG-results} and Fig.~\ref{chemistry}). In Arp~220, analyzed in detail by Graci{\'a}-Carpio et al.\@ (in preparation) with the additional input provided by the H$^{13}$CN and H$^{13}$CO$^{+}$ rotational lines, it is estimated that [HCN]$/$[HCO$^{+}$] $\geq 30$, in sharp contrast to the case of Arp~299\,A where we estimate that [HCN]$/$[HCO$^{+}$] $\sim 2$. IRAS\,17208--0014 represents an intermediate case where [HCN]$/$[HCO$^{+}$] $\sim 10$.

As discussed in Sec.~\ref{Introduction}, different scenarios may account for HCN being overabundant in extreme ULIRGs. But more relevant to the discussion in this paper is the fact that high HCN abundances call for a lower $L'_{\rm HCN(1-0)}$-to-$M_{\rm dense}$ conversion factor ($X_{\rm HCN}$), particularly at high $L_{\rm FIR}$. If the emission of HCN(1--0) is optically thin (a very unlikely scenario), the value of $X_{\rm HCN}$ should scale down as 1$/$[HCN]. It must be emphasized that even in the case of optically thick lines (a more realistic scenario, favored by our LVG analysis), a significant change in $X_{\rm HCN}$, due to the different properties of the emitting gas, cannot be excluded either \citep[see also discussion in][]{Papadopoulos07a, Papadopoulos07b}. This change can be quantitatively evaluated from the LVG results reported above. If we assume that the HCN(1--0) line comes from an ensemble of virialized, non-shadowing molecular clumps with optically thick emission, $X_{\rm HCN}$ can be estimated from the expression first derived by \citet{Dickman86}, as: 

\begin{equation}
  X_{\rm HCN}\sim 2.1\times \frac{n_{\rm H_{2}}^{1/2}}{T_{\rm HCN}} 
\label{Xhcn}
\end{equation}

\noindent where $n_{\rm H_{2}}$ and $T_{\rm HCN}$ represent respectively the density and brightness temperature derived from the LVG solution for the HCN(1--0) line. Based on the output from the best fit models (summarized in Table~\ref{LVG-results}), we estimate that the $X_{\rm HCN}$ value is $\sim$4.5 times lower at $L_{\rm FIR} \sim 10^{12}\,L_{\sun}$ compared to the value typically found at $L_{\rm FIR} \sim 10^{11}\,L_{\sun}$. While a better quantitative assessment of how $X_{\rm HCN}$ changes on average with $L_{\rm FIR}$ would require multi-phase radiative transfer models, a direct implication of this finding is that the reported trend in the $L_{\rm FIR}/L'_{\rm HCN}$ ratio as a function of $L_{\rm FIR}$, discussed in Sect.~\ref{SFE}, may be hiding a potentially more dramatic change of the SFE$_{\rm dense}$.

\subsection{Two-phase Large Velocity Gradient (LVG) models\label{two-phase}}

It is noteworthy that the simplest one-phase LVG approach developed in Sect.~\ref{one-phase} is already able to fit the three line ratios in most of the galaxies of our sample with a satisfactorily low $\chi^{2}$ figure ($<$3), except for three sources: in NGC~6240, NGC~7469 and NGC~7771, the derived $\chi^{2}$ is too large ($\gg$3). More sophisticated models, accounting for the multi-phase nature of molecular gas, can be used to try to significantly improve the fit to the three molecular line ratios observed in the galaxies of our sample in general, and in the three conflicting sources referred to above in particular. In addition, with the help of multi-phase models, we can explore if the main conclusions of Sect.~\ref{one-phase}, favoring a significant change of $X_{\rm HCN}$ with $L_{\rm FIR}$ are a mere artefact of one-phase LVG schemes. One-phase models may not properly take into account that the critical densities of the rotational lines of HCN and HCO$^{+}$ for the same J level differ by a factor $\sim$6. The latter would favor that the emission of these two lines may come, on average, from two distinct phases characterized by different densities. 

The main drawback of multi-phase models is that the number of free parameters to explore a priori is exceedingly large, however. The solutions are usually degenerate and the relevant parameters cannot be well constrained if the available number of data points to fit is reduced compared to the degrees of freedom. In our case, the constraints are given by the three line ratios independently measured in each galaxy, and also, by the the upper limit on the size of the source derived from CO interferometer maps ($\theta_{\rm CO}$ in Table~\ref{Table1}).

In order to minimize the number of free parameters, we have reduced the number of phases to two: a hot ($T_{\rm k} = 80$\,K) and \emph{dense} ($n_{\rm H_{2}} = 10^{5}$\,cm$^{-3}$) phase, which would be more directly associated with massive star formation, and a cold ($T_{\rm k} = 25$\,K) and comparatively more \emph{diffuse} phase ($n_{\rm H_{2}} = 10^{4}$\,cm$^{-3}$), which would account for a quiescent reservoir of molecular gas. This dichotomy in $T_{\rm k}$ reflects the two-temperature fit of the average IR SED of the galaxies in our sample. Furthermore, the density range explored by the \emph{dense} and \emph{diffuse} phases purposely encompasses the values of the effective critical densities of HCN(1--0) and HCO$^{+}$(1--0) respectively. The [HCN]$/$[HCO$^{+}$] abundance ratios are also fixed for the two phases as follows. We adopt a canonical value of [HCN]$/$[HCO$^{+}$] $ = 1$ for the \emph{diffuse} phase, typical of quiescent molecular dark clouds in our Galaxy \citep{Ohishi92,Pratap97,Dickens00}. In contrast, we fix [HCN]$/$[HCO$^{+}$] $= 10^{3}$ in the \emph{dense} phase, a value typical of galactic star forming regions where hot-core like chemistry can develop \citep{Johansson84,Blake87}. This choice of physical and chemical parameters for the two phases is expected to alleviate the need of enhancing the HCN \emph{global} abundance\footnote{We mean by \emph{global} HCN abundance, the average value derived from the two-phase model.} with respect to that of HCO$^{+}$ in some of the sources, a result found in Sect.~\ref{one-phase}. The robustness of this result and its implication for the constancy of $X_{\rm HCN}$ can thus be tested. A side effect of purposely fixing some of the parameters of the fit, for the reasons explained above, is that the resulting $\chi^{2}$ may be higher, i.e, formally worse, in the two-phase fit.

\begin{table*}
\caption{One-phase and two-phase LVG best fit solutions for the sample of LIRGs and ULIRGs detected in the J=1--0 and 3--2 lines of HCN and HCO$^{+}$. $\chi^{2}$ values of the fits are given in Col.~2. The predicted line ratios are in Cols.~3 to 5, followed by the [HCN]/[HCO$^{+}$] abundance ratio (Col.~6), the $L'_{\rm HCN(1-0)}$-to-$M_{\rm dense}$ conversion factor (Col.~7) and the estimated molecular source size at FWHM (Col.~8). For the two-phase LVG model, the relative filling factor of the dense phase is also given (Col.~9). Galaxies are ordered with decreasing $L_{\rm FIR}$. $L' = \rm K\,km\,s^{-1}\,pc^{2}$. $^{\mathrm{a}}$ Global values computed according to Equations~\ref{global-abundance} and \ref{global-Xhcn}.}
\label{LVG-results}
\centering
\begin{tabular}{@{}l r ccc rrrr@{}}
\multicolumn{4}{@{}l}{One-phase LVG model} & & & & & \\
\noalign{\smallskip}
\cline{1-8}
\noalign{\smallskip}
 Source & \multicolumn{1}{c}{$\chi^{2}$} & $\frac{\rm HCN(1-0)}{\rm HCO^{+}(1-0)}$ & $\frac{\rm HCN(3-2)}{\rm HCO^{+}(3-2)}$ & $\frac{\rm HCO^{+}(3-2)}{\rm HCO^{+}(1-0)}$ & \multicolumn{1}{c}{$\frac{\rm [HCN]}{\rm [HCO^{+}]}$} & \multicolumn{1}{c}{$X_{\rm HCN}$}         & \multicolumn{1}{c}{${\theta_{\rm s}}$} &   \\ 
        &                                &                                         &                                         &                                             &                                                       & \multicolumn{1}{c}{($M_{\sun}\,L'^{-1}$)} & \multicolumn{1}{c}{(\arcsec)}        &   \\ 
\noalign{\smallskip}
\cline{1-8}
\noalign{\smallskip}
 IRAS\,17208--0014 &  0.0\hspace{0.3cm} & 1.49 & 1.58 & 0.37 & 10.5\hspace{0.15cm} &  45\hspace{0.5cm} & 0.67 &       \\ 
 Mrk 231           &  2.7\hspace{0.3cm} & 1.04 & 1.08 & 0.33 &  6.8\hspace{0.15cm} &  26\hspace{0.5cm} & 0.85 &       \\ 
 Arp 220           &  1.3\hspace{0.3cm} & 2.32 & 3.92 & 0.23 & 28.0\hspace{0.15cm} &  22\hspace{0.5cm} & 1.24 &       \\ 
 Mrk 273           &  0.0\hspace{0.3cm} & 0.99 & 1.04 & 0.45 &  6.6\hspace{0.15cm} &  17\hspace{0.5cm} & 0.39 &       \\ 
 NGC 6240          & 33.3\hspace{0.3cm} & 0.55 & 0.38 & 0.20 &  2.5\hspace{0.15cm} & 127\hspace{0.5cm} & 1.69 &       \\ 
 Arp 193           &  0.0\hspace{0.3cm} & 0.63 & 0.37 & 0.34 &  2.2\hspace{0.15cm} & 130\hspace{0.5cm} & 1.14 &       \\ 
 Arp 299\,A        &  0.0\hspace{0.3cm} & 0.53 & 0.33 & 0.28 &  2.0\hspace{0.15cm} & 127\hspace{0.5cm} & 1.56 &       \\ 
 NGC 7469          &  6.5\hspace{0.3cm} & 0.69 & 0.54 & 0.17 &  3.5\hspace{0.15cm} & 147\hspace{0.5cm} & 2.09 &       \\ 
 Mrk 331           &  0.0\hspace{0.3cm} & 0.79 & 0.55 & 0.26 &  3.4\hspace{0.15cm} & 123\hspace{0.5cm} & 1.22 &       \\
 NGC 7771          & 12.1\hspace{0.3cm} & 1.01 & 0.85 & 0.18 &  5.4\hspace{0.15cm} & 129\hspace{0.5cm} & 2.12 &       \\ 
\cline{1-8}
\noalign{\medskip}
\noalign{\medskip}
\multicolumn{4}{@{}l}{Two phase LVG model} & & & & & \\
\noalign{\smallskip}
\hline 
\noalign{\smallskip}
 Source & \multicolumn{1}{c}{$\chi^{2}$} & $\frac{\rm HCN(1-0)}{\rm HCO^{+}(1-0)}$ & $\frac{\rm HCN(3-2)}{\rm HCO^{+}(3-2)}$ & $\frac{\rm HCO^{+}(3-2)}{\rm HCO^{+}(1-0)}$ & \multicolumn{1}{r}{$\frac{\rm [HCN]}{\rm [HCO^{+}]}^{\rm a}$} & \multicolumn{1}{c}{${X_{\rm HCN}}^{\rm a}$}         & \multicolumn{1}{c}{${\theta_{\rm s}}$} & \multicolumn{1}{c}{${f}$}  \\ 
        &                                &                                         &                                         &                                             &                                                               & \multicolumn{1}{c}{($M_{\sun}\,L'^{-1}$)} & \multicolumn{1}{c}{(\arcsec)}        & \multicolumn{1}{c}{(\%)} \\ 
\noalign{\smallskip}
\hline 
\noalign{\smallskip}
 IRAS\,17208--0014 &  4.8\hspace{0.3cm} & 1.16 & 1.79 & 0.42 &  3.1\hspace{0.15cm} &  23\hspace{0.5cm} & 0.60 & 17.4\hspace{0.3cm} \\
 Mrk 231           & 11.8\hspace{0.3cm} & 0.84 & 1.24 & 0.37 &  2.2\hspace{0.15cm} &  31\hspace{0.5cm} & 0.85 & 10.5\hspace{0.3cm} \\
 Arp 220           & 17.9\hspace{0.3cm} & 1.77 & 4.46 & 0.25 &  4.2\hspace{0.15cm} &  35\hspace{0.5cm} & 1.28 & 24.2\hspace{0.3cm} \\
 Mrk 273           &  0.4\hspace{0.3cm} & 0.90 & 1.15 & 0.46 &  2.4\hspace{0.15cm} &  23\hspace{0.5cm} & 0.42 & 12.0\hspace{0.3cm} \\
 NGC 6240          &  0.9\hspace{0.3cm} & 0.57 & 1.97 & 0.19 & 49.6\hspace{0.15cm} & 132\hspace{0.5cm} & 1.73 &  1.6\hspace{0.3cm} \\
 Arp 193           &  4.9\hspace{0.3cm} & 0.48 & 0.43 & 0.36 &  1.3\hspace{0.15cm} &  44\hspace{0.5cm} & 0.94 &  2.6\hspace{0.3cm} \\
 Arp 299\,A        &  7.7\hspace{0.3cm} & 0.47 & 0.47 & 0.31 &  1.3\hspace{0.15cm} &  56\hspace{0.5cm} & 1.27 &  2.7\hspace{0.3cm} \\
 NGC 7469          &  1.2\hspace{0.3cm} & 0.67 & 1.61 & 0.18 &  1.7\hspace{0.15cm} &  88\hspace{0.5cm} & 1.53 &  7.0\hspace{0.3cm} \\
 Mrk 331           & 10.1\hspace{0.3cm} & 0.64 & 0.84 & 0.34 &  1.7\hspace{0.15cm} &  40\hspace{0.5cm} & 0.87 &  6.3\hspace{0.3cm} \\
 NGC 7771          & 11.5\hspace{0.3cm} & 0.85 & 2.63 & 0.15 &  2.0\hspace{0.15cm} & 103\hspace{0.5cm} & 1.96 &  9.4\hspace{0.3cm} \\
\hline 
\end{tabular}
\end{table*}

In the fitting procedure, the relative filling factor of the two phases ($f$ for the dense gas and 1$-f$ for the diffuse phase) is allowed to vary. We also allow to freely fit the column density of HCN in the two phases with the constrain that [HCN]$^{\rm dense} \geq $ [HCN]$^{\rm diffuse}$, in order to be consistent with the values derived in our Galaxy for cold dark clouds and star forming regions. As in Sect.~\ref{one-phase}, we only keep the LVG solutions with $\eta_{\rm fill}$ values for HCN and HCO$^{+}$ lines that are below those derived for CO. Table~\ref{LVG-results} summarizes the results of the two-phase model fitting. To ease the comparison with one-phase models, we have derived the \emph{global} [HCN]$/$[HCO$^{+}$] abundance ratios and $X_{\rm HCN}$ conversion factors characterizing the best fit two-phase solutions. Global estimates are obtained by averaging the contributions from the two phases, as follows:

\begin{equation}
  \frac{\rm [HCN]}{\rm [HCO^{+}]}=\frac{f N_{\rm HCN}^{\rm dense} + (1-f) N_{\rm HCN}^{\rm diffuse}}
{f N_{\rm HCO^{+}}^{\rm dense} + (1-f) N_{\rm HCO^{+}}^{\rm diffuse}}
\label{global-abundance}
\end{equation}

\begin{equation}
  X_{\rm HCN}=\frac{f T_{\rm HCN}^{\rm dense} X_{\rm HCN}^{\rm dense} + (1-f) T^{\rm diffuse}_{\rm HCN} X_{\rm HCN}^{\rm diffuse}}
{f T_{\rm HCN}^{\rm dense} + (1-f) T^{\rm diffuse}_{\rm HCN}} 
\label{global-Xhcn}
\end{equation}

In the first equation, $N_{\rm HCN}^{\rm dense}$ and $N_{\rm HCN}^{\rm diffuse}$ represent the column densities of HCN derived for the dense and the diffuse phase model clouds, respectively. Similar symbols are used for HCO$^{+}$. In the second equation, $X_{\rm HCN}^{\rm dense}$ and $X_{\rm HCN}^{\rm diffuse}$ are the $X_{\rm HCN}$ factors calculated for the dense and the diffuse phase, respectively, according to Equation~\ref{Xhcn}. The brightness temperatures of HCN(1--0) in the dense ($T_{\rm HCN}^{\rm dense}$) and in the diffuse phases ($T^{\rm diffuse}_{\rm HCN}$) are used as weighting factors in each phase.

First, as a bottom line conclusion, we see that the goodness of the fit of the two-phase models, measured by $\chi^{2}$, is worse compared to that achieved in the one-phase models. This applies to all sources, with the exception of NGC~6240 and NGC~7771 (see Table~\ref{LVG-results}). Although with a large scatter, the best-fit models indicate that the filling factor ($f$) of the dense phase increases with $L_{\rm FIR}$. This result can be taken as an evidence that the average density of molecular gas increases with $L_{\rm FIR}$. In addition, we find no significant difference in the [HCN]$/$[HCO$^{+}$] abundance ratios, estimated by Equation~\ref{global-abundance}, between ULIRGs and LIRGs, i.e., we eliminate the need of enhancing the \emph{global} abundance of HCN with respect to that of HCO$^{+}$ at high $L_{\rm FIR}$, in contrast with the result found in Sect.~\ref{one-phase}. NGC~6240 is, on this respect, also an exception to the rule. The increasing trend of $f$ with $L_{\rm FIR}$ indicated by the best-fit solutions seems to neutralize the need of a similar trend of [HCN]$/$[HCO$^{+}$] with $L_{\rm FIR}$ in the two-phase model. However, we find that the conversion factor $X_{\rm HCN}$, derived from Equation~\ref{global-Xhcn}, is lower at high $L_{\rm FIR}$, a conclusion similar to that found in one-phase models. In particular, $X_{\rm HCN}$ is $\sim$2.5 times lower at $L_{\rm FIR} \sim 10^{12}\,L_{\sun}$ compared to the value typically found at $L_{\rm FIR} \sim 10^{11}\,L_{\sun}$.

\section{Conclusions\label{Discussion}}

In this paper we present observational evidence that the $L_{\rm FIR}/L_{\rm HCN(1-0)}$ ratio, taken as proxy for SFE$_{\rm dense}$, is a factor $\sim$2--3 higher in galaxies categorized as IR luminous ($L_{\rm FIR} > 10^{11}\,L_{\sun}$) compared to normal galaxies. Local universe LIRGs and ULIRGs populate a region in the SFE$_{\rm dense}$ diagram that lies between those occupied by normal and high-$z$ IR luminous galaxies. The reported trend in the SFE$_{\rm dense}$ derived from HCN data implies that there is a statistically significant turn upward in the KS law, $\Sigma_{\rm SFR} \propto \Sigma_{\rm dense}^{N}$, at high $L_{\rm FIR}$: $N$ changes from $\sim$0.80--0.95 (for $L_{\rm FIR} < 10^{11}\,L_{\sun}$) to $\sim$1.1--1.2 (for $L_{\rm FIR} > 10^{11}\,L_{\sun}$). Furthermore, our multiline analysis of HCN and HCO$^{+}$ data indicates that $X_{\rm HCN}$ is $\sim$3 times lower at high $L_{\rm FIR}$. This latter finding reinforces a scenario where the SFE$_{\rm dense}$ may be up to an order of magnitude higher in extreme IR luminous galaxies than in normal galaxies. 

Based on the model developed by \citet{Krumholz05}, in which the SFR density ($\rho_{\rm SFR}$) scales with the average density of the gas as $\sim\overline{\rho}_{\rm gas}^{\ 1.5}$, \citet{Krumholz07b} conclude that the power index of KS laws determined directly from observations would change depending on how the effective critical density of the tracer used to probe the star forming gas compares to the average density of the gas itself. In the particular case of HCN(1--0) data, only in galaxies where $\overline{\rho}_{\rm gas}$ exceeds a few $10^{4}\,\rm cm^{-3}$ (i.e., the effective critical density of the HCN J=1--0 line), we would start recovering the expected superlinear behavior of the \emph{universal} KS law derived by \citet{Krumholz05}. Our results, hinting at a break in the KS power law derived from HCN data around $L_{\rm FIR} = 10^{11}\,L_{\sun}$, are in qualitative agreement with this picture. However, it remains to be proved if the one order of magnitude increase in the SFE$_{\rm dense}$ over 3 decades in $L_{\rm FIR}$ derived from our data can be fitted by a \emph{universal} KS law with a corresponding increase of $\overline{\rho}_{\rm gas}$ \citep[e.g., see Fig.~2 of][]{Krumholz07b}. Based on our two-phase LVG fits of LIRGs and ULIRGs molecular data, we do find tantalizing evidence of higher $\overline{\rho}_{\rm gas}$ at higher $L_{\rm FIR}$. Enlarging the number of detected molecules and transitions will be a key to better constraining $\overline{\rho}_{\rm gas}$. 

Alternatively, the reported increase of the SFE$_{\rm dense}$ at high $L_{\rm FIR}$ may be due to the fact that star formation processes deviate from the KS recipes followed by normal galaxies. The high frequency of interactions can create a pressure enhanced environment in the ISM of LIRGs and ULIRGs that may specifically elevate SFE$_{\rm dense}$ in these galaxies. On the other hand, an extreme ISM environment, characterized by high densities and temperatures, may also create a top-heavy Initial Stellar Mass Function (IMF) in luminous IR galaxies. This scenario has been suggested to apply to the nuclei of starbursts and to the center of the Galaxy \citep[e.g.,][]{Klessen07}. With this assumption, the conversion factor between SFR and $L_{\rm FIR}$, mostly IMF-dependent, should be lowered at high $L_{\rm FIR}$, instead of being kept constant as supposed in Sect.~\ref{SFE}. The SFE$_{\rm dense}$ trend derived above may thus be reflecting a change of the IMF rather than a true variation of the SFE.

Finally a significant contribution from an AGN source to $L_{\rm FIR}$ cannot be excluded, especially at the high luminosity end (cf. \citetalias{Gracia-Carpio06}). This would produce a trend in the SFE$_{\rm dense}$ similar to the one reported above but due to the AGN contamination being more severe at high $L_{\rm FIR}$. In this context, the discovery by \citet{Downes07} of an embedded AGN source in the West nucleus of Arp~220 is noteworthy: in this ULIRG, considered to date as a prototype of a starburst dominated IR galaxy \citep[e.g.,][]{Genzel98}, \citet{Downes07} claim that $\sim$75\% of the total IR luminosity of the galaxy may come from the accretion disk of the AGN. The fact that we have also derived in Arp~220 the most extreme [HCN]$/$[HCO$^{+}$] abundance ratio among all the galaxies of our sample, points to a causal link between HCN overabundance and AGN-driven chemistry. However, it is still a mostly controversial issue whether X-ray chemistry is able to enhance [HCN]$/$[HCO$^{+}$] abundance ratios in AGNs to the level required by observations \citep{Usero04,Meijerink05,Garcia-Burillo06,Meijerink07}. Other related processes involving the evaporation of dust grain mantles and high temperature gas-phase reactions, likely efficient in the dense and hot molecular environments typical of extreme starbursts, and probably also of AGNs, may be responsible of shaping the chemistry of the molecular gas in IR luminous galaxies \citep[\citetalias{Gracia-Carpio06};][]{Lahuis07}.
  
This paper shows how multiline analysis of molecular tracers of dense gas can contribute to study the star formation rate and star formation efficiency laws in starbursting galaxies, providing more elaborated tools to determine whether the SFE is enhanced or constant in the most extreme starbursts found in the local and the high-$z$ universe.

\begin{acknowledgements}
We thank the IRAM staff at the 30m telescope for their support during the observations. We thank Luis Colina for his comments on a preliminary version of the paper. This work has been partially supported by the Spanish MEC and Feder funds under grant ESP2003-04957.
\end{acknowledgements}

\bibliographystyle{aa} 

\bibliography{main.bib}

\end{document}